\documentclass[prd,10pt,twocolumn,tightenlines,superscriptaddress,numerical,notitlepage,showpacs,amsmath,amssymb,amsfonts,aps,nofootinbib,longbibliography,floatfix]{revtex4-1}

\usepackage{combelow}
\usepackage{dsfont}
\usepackage{amsmath,amssymb,mathtools,bm,bbm,braket,slashed}
\usepackage{tikz}
\usepackage{cancel}
\usepackage[caption=false]{subfig}

\usepackage{xcolor}
\usepackage{hyperref}
\hypersetup{colorlinks=true, citecolor=blue, urlcolor=blue, citecolor=red}

\definecolor{purple}{rgb}{0.8,0,0.6}

\definecolor{battleshipgrey}{rgb}{0.2, 0.52, 0.51}
\definecolor{darkgreen}{rgb}{0.12, 0.5, 0.17}

    \newcommand{\beqn}{\begin{eqnarray}}
    \newcommand{\eeqn}{\end{eqnarray}}
    \newcommand{\beqs}{\begin{subequations}}
    \newcommand{\eeqs}{\end{subequations}\\[-2mm]\noindent}

\begin{document}

\title{Chiral restoration temperature at finite spin density in QCD}

\author{Pracheta Singha}
%\thanks{Corresponding author: victor.ambrus@e-uvt.ro.}
\affiliation{Department of Physics, West University of Timi\cb{s}oara,  Bd.~Vasile P\^arvan 4, Timi\cb{s}oara 300223, Romania}

\author{Victor E. Ambru\cb{s}}
\affiliation{Department of Physics, West University of Timi\cb{s}oara, Bd.~Vasile P\^arvan 4, Timi\cb{s}oara 300223, Romania}

\author{Sergiu Busuioc}
\affiliation{Department of Physics, West University of Timi\cb{s}oara,  Bd.~Vasile P\^arvan 4, Timi\cb{s}oara 300223, Romania}

\author{Aritra Bandyopadhyay}
\affiliation{Department of Physics, West University of Timi\cb{s}oara,  Bd.~Vasile P\^arvan 4, Timi\cb{s}oara 300223, Romania}

\author{Maxim N. Chernodub}
\affiliation{Institut Denis Poisson, CNRS UMR 7013, Universit\'e de Tours, Universit\'e d'Orl\'eans,
Parc de Grandmont, Tours, 37200, France}
\affiliation{Department of Physics, West University of Timi\cb{s}oara, Bd.~Vasile P\^arvan 4, Timi\cb{s}oara 300223, Romania}

\date{\today}

\begin{abstract}
We investigate the impact of a uniform spin density on the critical temperature of the chiral phase transition in finite-temperature QCD in the scope of the linear sigma model. We demonstrate that at a finite spin potential $\mu_\Sigma$, corresponding to a finite spin density, the predictive power of the model is challenged by an ambiguity associated with a contribution of the vacuum renormalization term to the free energy. Eliminating the regularization freedom through comparison with recent low-$\mu_{\Sigma}$ lattice data, we extend the phase diagram of QCD at finite spin density to regions inaccessible to lattice simulations. We show that, as the spin potential increases, the temperature of the chiral crossover transition diminishes and the chiral crossover turns into a first-order transition at a second-order critical end-point $(T,\mu_\Sigma)_\mathrm{CEP}\simeq (0.142,0.098)$ GeV. With increasing spin potential, the critical temperature touches the zero-temperature axis at $\mu_{\Sigma} = 0.310$ GeV, implying that the chiral symmetry is restored at higher potentials at any temperature.
\end{abstract}

\maketitle

\section{Introduction}
\label{sec:Introduction}

Spin polarization phenomena in the quark--gluon plasma (QGP) have recently attracted significant attention~\cite{Becattini:2022zvf}. Observations of global and local spin polarizations in non-central heavy-ion collisions~\cite{STAR:2017ckg,STAR:2023eck} reveal the transfer of angular momentum from colliding nuclei to hadronic matter and provide sensitive probes of the thermodynamic, topological, and electromagnetic properties of QCD matter~\cite{Huang:2020dtn,Fukushima:2018grm,Becattini:2024uha}.

Measurements of the global polarization of $\Lambda$ hyperons and of local spin polarization---including the spin alignment of various hadrons---suggest that, at a certain stage of its evolution, the vortical QGP develops a finite spin density of quarks. A precise analytical formulation of spin degrees of freedom,
however, remains subtle due to the so-called ``pseudogauge symmetry ambiguity''~\cite{Becattini:2018duy,Speranza:2020ilk,Fukushima:2020ucl}, which reflects the
non-uniqueness of a local definition of the spin operator. This ambiguity can potentially be resolved within interacting field theories, as recently demonstrated in Ref.~\cite{Buzzegoli:2024mra}, through the canonical spin formulation of quarks in an effective infrared model of QCD.

Based on this idea, in our paper we explore the effect of finite spin quark density on the QGP thermodynamics and chiral transition properties within the Linear sigma model coupled to quarks~\cite{GellMann1960}, LSM$_q$, using the canonical formulation of the spin operator. The spin imbalance is described by the spin potential, $\mu_{\Sigma}$, which plays the role of a thermodynamic quantity conjugated to the spin density and also enters the hydrodynamical description of the system~\cite{Montenegro:2017rbu, Florkowski:2019gio, Gallegos2021, Ambrus:2022yzz, Abboud:2025qtg, Becattini:2025oyi}. A similar incorporation of spin density has been recently employed in first principle lattice simulation with two dynamical light quarks~\cite{Braguta:2025ddq}, where it was shown that a finite spin potential leads to a decrease of both chiral restoration and deconfinement temperatures in the limit of the physical quark masses.

It is worth mentioning that the spin polarization, being a part of the total angular momentum of the system, is not a separately conserved quantity because the angular momentum can be shared between the orbital and spin contributions. Therefore, the spin potential should be distinguished from genuine chemical potentials associated, for example, with the conserved electric, baryonic or isospin charges. Still, the lattice simulations of the QCD phase diagram at finite spin potential have demonstrated the excellent continuum scaling of the physical quantities~\cite{Braguta:2025ddq}. The absence of infrared and ultraviolet artifacts in these simulations suggests that the implementation of the spin potential is free from pathologies that could be associated, for example, with the spin non-conservation and the pseudogauge freedom. Therefore, the first-principle lattice simulations suggest that the canonical spin potential for quark degrees of freedom has a physical sense in QCD.

However, similar to the baryon chemical potential ($\mu_B$), due to the sign problem in the lattice implementation of real $\mu_\Sigma$, the lattice results reported in Ref.~\cite{Braguta:2025ddq} were limited to low spin potentials, effectively restricting the exploration of regimes with substantial spin potential. In our paper, we study the influence of the spin potential on the thermodynamics of QCD within the effective model framework which, theoretically, has no such limitation. Thus, the aim of the current project is twofold: (i) to match the predictions of the model for the chiral transition line to the lattice data at small values of the spin potential; and (ii) to explore the region of arbitrary values of the spin potential, which is inaccessible to lattice simulations.

One advantage of using the LSM$_q$ model over other effective QCD models is that it is renormalizable. In this work, we explore the effect of vacuum renormalisation on the model thermodynamics. We find that incorporating $\mu_\Sigma$-dependent finite counter-terms in the vacuum free energy turns out to be crucial for obtaining a qualitative agreement with the lattice results of Ref.~\cite{Braguta:2025ddq}. Within this setup, we investigate the influence of the spin potential on the phase diagram of QCD. We also study the behavior of spin density as a function of the thermodynamic parameters $T$ and $\mu_\Sigma$.

We mention that the nonconservation of the canonical spin tensor has profound implications, as will be explicitly demonstrated through our analysis. First of all, it does not commute with the Dirac Hamiltonian. This gives rise to a nonlinear dispersion relation that leads to a $\mu_\Sigma$-dependent zero-point free energy. Second, unlike the baryon chemical potential, the spin potential has no Fermi level. Third, at large temperatures, the thermal part of the quark free energy becomes independent of $\mu_\Sigma$ and the spin expectation value is due solely to the renormalized meson potential. None of the above (unnatural) features are present in the case when the polarization is induced via the conserved helicity operator, considered in our previous work \cite{Chernodub:2020yaf}. It would be interesting to test this hypothesis with other conserved spin operators, such as the Pryce operator introduced in Ref.~\cite{Pryce:1948} and revisited recently in Refs.~\cite{Bauke_2014,Cotaescu:2022ntp}.

The structure of this paper is as follows. Section~\ref{sec:Mod} introduces the details of the model. Section~\ref{subsec:Mes} is devoted to the meson sector, while Section~\ref{subsec:quark} addresses the quark sector, where the finite spin density term is incorporated into the quark Lagrangian and the corresponding dispersion relation is derived. The thermodynamics of the model is discussed in Section~\ref{Subsec:thermodynamics}. In Section~\ref{sec:vac_ren}, we present a detailed analysis of vacuum renormalization, which plays a central role in achieving a consistent description of QCD thermodynamics with finite spin density within the LSM$_q$ framework. Section~\ref{sec:joy} highlights the subtleties involved in solving the saddle-point equations with the modified dispersion relation in the presence of a spin potential. The calculation of the phase diagram and determination of the ambiguity of the renormalization procedure with the help of the lattice data are presented in Section~\ref{sec:res}. Our conclusions are summarized along with a brief discussion in Section~\ref{sec:conclusion}.
Four appendices are devoted to the discussion of dimensional regularization for vacuum renormalization (Appendix~\ref{app:DR}), the thermal quark contributions (Appendices~\ref{app:sad} and \ref{app:deltaFq}), and the connection between our renormalization scheme and the wave function renormalization approach of Ref.~\cite{Brandt:2025tkg} (Appendix~\ref{app:rg}).
Throughout this work, we adopt the Minkowski metric $g = {\rm diag}(1, -1, -1, -1)$.

\section{Model}
\label{sec:Mod}

The linear sigma model coupled to quarks~\cite{GellMann1960} describes interacting meson and quark degrees of freedom. For the 2-flavor system considered in this work, the quark sector is given by the isospin doublet $\psi=(u,d)^T$, where $u$ and $d$ denote the up and down quark spinors, respectively. In the meson sector, we consider the scalar $\sigma$ meson and the isotriplet pseudoscalar pion mesons $\vec{\pi}=(\pi^+,\pi^-,\pi^0)$. The model Lagrangian can be written as follows:
\begin{align}
\mathcal{L}=\mathcal{L}_\mathcal{M} + \mathcal{L}_q\,,
\label{eq_L_LSMq}
\end{align}
where $\mathcal{L}_\mathcal{M}$ and $\mathcal{L}_q$ correspond to the meson and the quark parts of the Lagrangian, respectively, which we will briefly discuss in the following.
\subsection{Meson sector}\label{subsec:Mes}
The meson sector incorporates the chiral properties of the model. In the QCD with finite quark masses, the chiral symmetry is explicitly broken, and the chiral phase transition manifests as a smooth crossover rather than a true phase transition. Consequently, the field $\sigma$ plays the role of the approximate order parameter associated with the chiral symmetry breaking: it takes a non-vanishing value in the low-temperature, chirally broken phase and gradually approaches zero as the temperature increases.

The meson part of the Lagrangian,
\begin{align}
    \mathcal{L}_\mathcal{M}=\frac{1}{2}\left(\partial_\mu \sigma\partial^\mu \sigma + \partial_\mu \vec{\pi} \cdot \partial^\mu \vec{\pi}\right)-V_\mathcal{M},
    \label{eq_L_meson}
\end{align}
incorporates, besides the kinetic terms for the meson fields, the phenomenological interaction potential:
\begin{align}
    V_\mathcal{M} =\frac{\lambda}{4}(\sigma^2 +\vec{\pi}^2 -v^2)^2 - h \sigma,
    \label{eq:pot_meson}
\end{align}
with $\lambda$, $v$ and $h$ being model parameters.

The breaking of the chiral symmetry in the vacuum is implemented at the level of the $\sigma$ meson, which acquires a non-vanishing vacuum expectation value. In contrast, the pion field retains a vanishing vacuum expectation value, $\langle {\vec \pi} \rangle = 0$, as it can always be removed by a global chiral rotation of the pion and sigma fields (see, for example, Ref.~\cite{Schechter:1971efy}). The first term in the potential~\eqref{eq:pot_meson} leads to a strong spontaneous breaking of the chiral symmetry, supplemented by weak explicit symmetry breaking due to the second term.

The potential~\eqref{eq:pot_meson} can be expanded over the small fluctuations $\delta\sigma$ and $\delta\pi$ of the meson fields as:
\begin{align}
    V_\mathcal{M} = V_0 + \frac{1}{2} m_\sigma^2\delta\sigma^2+\frac{1}{2}m_\pi^2\delta\vec{\pi}^2+\cdots\,,
\end{align}
which allows the meson masses to be computed as:
\begin{align}
    m_\sigma^2=\lambda(3\langle\sigma\rangle^2 -v^2)\,, \qquad m_\pi^2=\lambda(\langle\sigma\rangle^2-v^2)\,.
    \label{eq_mass_meson}
\end{align}
The masses of the charged and the neutral pions do not split due to the absence of electromagnetic interactions.

The parameter $h$ ensuring the explicit symmetry breaking is obtained by enforcing the PCAC relation $h=f_\pi m_\pi^2$, where $f_\pi = 93$ MeV is the pion decay constant and $m_\pi=138$ MeV is the pion mass in the vacuum. Demanding $\langle\sigma\rangle=f_\pi$~\cite{GellMann1960} and $m_\sigma=600$ MeV, the remaining model parameters $v$ and $\lambda$ are fixed through the relations $v^2=f_\pi^2-m_\pi^2/\lambda$ and $m_\sigma^2=2\lambda f_\pi^2+m_\pi^2$.

\subsection{Quark sector}
\label{subsec:quark}
We extend the usual quark sector of the linear sigma model coupled to quarks (LSM$_q$), which includes both the quark-meson interaction and the kinetic term for quark fields, to account for a non-vanishing spin density:
\begin{equation}
{\mathcal L}_q = \overline{\psi} \left[i {\slashed \partial}
- g (\sigma + i \gamma_5 \vec{\tau} \cdot {\vec \pi} ) \right]\psi +
\delta_\Sigma \mathcal{L}_q.
\label{eq:Lag_quark_wspin}
\end{equation}
The interaction term involving the $\sigma$ and $\vec{\pi}$ fields enables dynamical mass generation for the quarks through spontaneous chiral symmetry breaking, with $\langle\sigma\rangle \neq 0$ leading to a constituent quark mass $m_q = g \langle\sigma\rangle$. Taking $m_q = 307$ MeV yields a Yukawa coupling $g \simeq 3.3$.

We express the additional term $\delta_\Sigma \mathcal{L}_q$ as \cite{Braguta:2025ddq}
\begin{align}
    \delta_\Sigma {\mathcal L}_q = \mu_{\alpha,\mu\nu} {\cal S}^{\alpha,\mu\nu} \,,
    \label{eq:Lag_quark_sp}
\end{align}
where the canonical spin part of the angular momentum density for spin-1/2 Dirac fermions is given by
\begin{align}
    {\cal S}^{\alpha,\mu\nu} = \frac{1}{2} \bar{\psi} \bigl\{\gamma^\alpha, \Sigma^{\mu\nu}\bigr\}\psi \,,
    \qquad
    \Sigma^{\mu\nu} = \frac{i}{4} \bigl[\gamma^\mu, \gamma^\nu\bigr]\,.
    \label{eq_spin_operator}
\end{align}
The associated background relativistic spin field $\mu_{\alpha,\mu\nu}$ introduces a finite spin density and a finite spin current in the system. For the study of polarization, we consider the case when only the rotation part of the spin potential is non-vanishing. Defining $\mu_{\alpha\beta} = -\frac{1}{2} \varepsilon_{0\alpha\beta k} \mu_\Sigma^k$, with $\varepsilon^{0123} = -\varepsilon_{0123} = 1$, we arrive at
\begin{equation}
 \delta_\Sigma \mathcal{L}_q = \mu_\Sigma \psi^\dagger (\mathbf{n}_\Sigma \cdot \boldsymbol{\Sigma}) \psi,
 \label{eq:dL_AV_gen}
\end{equation}
where we took $\mathbf{n}_\Sigma = \boldsymbol{\mu}_\Sigma / \mu_\Sigma$ to be the direction of the vector spin potential, with $\mu_\Sigma = \sqrt{\mu_{\Sigma;1}^2 +\mu_{\Sigma;2}^2 + \mu_{\Sigma;3}^2}$ being the three-norm of $\boldsymbol{\mu}_\Sigma$. We also used the convention $\Sigma^i = \frac{1}{2} \varepsilon^{0ijk} \Sigma_{jk}$, i.e. $\Sigma^3 = \Sigma^{12} = \frac{i}{2} \gamma^1 \gamma^2$, with the other components given by circular permutations. The spin term~\eqref{eq:Lag_quark_sp} in the Lagrangian~\eqref{eq:Lag_quark_wspin} does not break explicitly the global chiral symmetry,  as it remains invariant under the transformations $\psi \to e^{i \boldsymbol{\tau}\cdot\boldsymbol{\omega}_5 \gamma_5} \psi$ and ${\bar\psi} \to {\bar\psi} e^{i\boldsymbol{\tau}\cdot \boldsymbol{\omega}_5 \gamma_5}$ with a coordinate-independent, real-valued parameter $\boldsymbol{\omega}_5$.

In the mean-field approximation, the meson fields are replaced by their (homogeneous) expectation values in the quark Lagrangian \eqref{eq:Lag_quark_wspin}. Taking into account that $\langle {\vec \pi} \rangle = 0$, the Dirac Hamiltonian operator reads as follows:
\begin{equation}
    H_q = -i\gamma^0 \boldsymbol{\gamma}\cdot \boldsymbol{\nabla} + \gamma^0 g\sigma - \mu_\Sigma (\mathbf{n}_\Sigma \cdot \boldsymbol{\Sigma}).
\end{equation}
To study the energy branches supported by the system, we construct the eigensystem of $H_q$ at the level of Fourier modes $e^{i \mathbf{p} \cdot \mathbf{x}} \psi_{\varsigma}^{(s)}(\mathbf{p})$,
\begin{equation}
 H_q(\mathbf{p}) \psi_{\varsigma}^{(s)}(\mathbf{p}) = \varsigma E_{\mathbf{p}}^{(s)} \psi_{\varsigma}^{(s)}(\mathbf{p}),
 \label{eq:H_eigen}
\end{equation}
with $\varsigma = \pm 1$ labeling particle ($+1$) and antiparticle ($-1$) modes, while $s = \pm 1/2$ denotes the polarization degree of freedom. Denoting $\psi_{\varsigma}^{(s)} = (\psi_{\varsigma;u}^{(s)}, \psi_{\varsigma;d}^{(s)})^T$, where $\psi_{\varsigma;u/d}^{(s)}$ represent upper/lower two-component spinors, we have
\begin{align}
 (\varsigma E^{(s)}_{\mathbf{p}} - g\sigma + \tfrac{1}{2} \mu_\Sigma \mathbf{n}_\Sigma \cdot \boldsymbol{\sigma}) \psi_{\varsigma;u}^{(s)} &= \boldsymbol{\sigma} \cdot \mathbf{p}\, \psi_{\varsigma;d}^{(s)}, \\
 (\varsigma E^{(s)}_{\mathbf{p}} + g\sigma + \tfrac{1}{2} \mu_\Sigma \mathbf{n}_\Sigma \cdot \boldsymbol{\sigma}) \psi_{\varsigma;d}^{(s)} &= \boldsymbol{\sigma} \cdot \mathbf{p} \,\psi_{\varsigma;u}^{(s)}.
\end{align}
Multiplying the top relation by $\boldsymbol{\sigma} \cdot \mathbf{p}$ and using $(\boldsymbol{\sigma} \cdot \mathbf{p})^2 = \mathbf{p}^2$, $\psi_{\varsigma;d}^{(s)}$ can be eliminated in favor of $\psi_{\varsigma;u}^{(s)}$, leading to
\begin{equation}
 (\mathbf{A} \cdot \boldsymbol{\sigma} - B) \psi_{\varsigma;u}^{(s)} = 0,
 \label{eq:dispersion_aux}
\end{equation}
where we denoted
\begin{align}
 \mathbf{A} &= \varepsilon^2 \mathbf{p} + \frac{1}{2} \mu_\Sigma^2 (\mathbf{p} \cdot \mathbf{n}_\Sigma) \mathbf{n}_\Sigma + i \mu_\Sigma g\sigma (\mathbf{p} \times \mathbf{n}_\Sigma), \nonumber\\
 B &= -\mu_\Sigma \varsigma E^{(s)}_{\mathbf{p}} (\mathbf{p} \cdot \mathbf{n}_\Sigma),
\end{align}
with $\varepsilon^2 = [E^{(s)}_{\mathbf{p}}]^2 - \mathbf{p}^2 - g^2 \sigma^2 - \mu_\Sigma^2 /4$. We now multiply Eq.~\eqref{eq:dispersion_aux} by $\mathbf{A} \cdot \boldsymbol{\sigma} + B$. Using the relation $(\mathbf{A} \cdot \boldsymbol{\sigma} - B) (\mathbf{A} \cdot \boldsymbol{\sigma} + B) = \mathbf{A}^2 - B^2$ for any C-numbers $\mathbf{A}$, $B$, non-trivial solutions require that $\mathbf{A}^2 = B^2$. After some simplifications, we arrive at
\begin{equation}
 \varepsilon^4 = \mu_\Sigma^2[(\mathbf{p} \cdot \mathbf{n}_\Sigma)^2 + g^2 \sigma^2].
\end{equation}
We employ the polarization parameter $s = \pm 1/2$ to keep track of the sign of the square root of the above expression, such that
\begin{equation}
 [E^{(s)}_{\mathbf{p}}]^2 = \mathbf{p}^2 + g^2\sigma^2 + s^2 \mu_\Sigma^2 - 2s\mu_\Sigma \sqrt{(\mathbf{p} \cdot \mathbf{n}_\Sigma)^2 + g^2 \sigma^2}.
\end{equation}
Taking another square root and demanding that $E^{(s)}_{\mathbf{p}} {>} 0$, we arrive at
\begin{align}
 E_{\mathbf{p}}^{(s)} &= \sqrt{p^2 + g^2 \sigma^2 + s^2\mu_\Sigma^2 - 2s\mu_\Sigma \sqrt{(\mathbf{p} \cdot \mathbf{n}_\Sigma)^2 + g^2 \sigma^2}} \nonumber\\
 &= \sqrt{(\mathbf{p} \times \mathbf{n}_\Sigma)^2 + [\sqrt{(\mathbf{p} \cdot \mathbf{n}_\Sigma)^2 + g^2\sigma^2} - s\mu_\Sigma]^2}.
 \label{eq:Energy_dispersion_mus_general}
\end{align}
Specializing the above result to the case when the spin polarization is directed along the $z$ axis, i.e. $\mathbf{n}_\Sigma = \mathbf{e}_z$, we obtain
\begin{align}
 E_{\mathbf{p}}^{(s)} & =  \sqrt{p^2 + g^2 \sigma^2 + s^2 \mu_\Sigma^2 - 2s \mu_\Sigma\sqrt{p_z^2 + g^2 \sigma^2} }  \nonumber\\
 & =  \sqrt{E_p^2 + s^2 \mu_\Sigma^2 - 2s \mu_\Sigma E_{p_z} },     \label{eq:Energy_dispersion_mus_final}
\end{align}
where $E_p = \sqrt{p^2 + g^2 \sigma^2}$ and $E_{p_z} = \sqrt{p_z^2 + g^2 \sigma^2}$ are, respectively, the total and longitudinal energies in the limit of vanishing spin potential. The very same spatially-anisotropic form of the dispersion relations also appears in condensed matter systems, such as in Weyl semimetals~\cite{Burkov2011} or in Dirac semimetals subjected to rotating electric fields~\cite{Ebihara:2015aca}. In addition, the double-square-root structure of the energy dispersion~\eqref{eq:Energy_dispersion_mus_final} emerges also in QCD in the presence of
chiral~\cite{Farias:2016let, Azeredo:2024sqc} and isospin~\cite{Avancini:2019ego, Lopes:2021tro, Ayala:2023cnt} chemical potentials.

It is instructive to clarify the physical meaning of the spin potential $\mu_\Sigma$. This quantity enters the Lagrangian, Eqs.~\eqref{eq:Lag_quark_wspin} and \eqref{eq:dL_AV_gen}, as a Lagrange multiplier in a way formally-analogous to a conventional chemical potential. However, there is a fundamental difference between the two quantities. A standard chemical potential quantifies the energy required to add or remove a particle carrying a conserved charge. For instance, the baryonic chemical potential $\mu_B$ corresponds to the energy cost of inserting a unit of baryonic charge (e.g., a proton) or, equivalently, removing a unit of negative baryonic charge (e.g., an antiproton) from the ensemble.

By contrast, the mentioned analogy fails for the spin potential: first, spin is not a conserved quantum number; and second, the quasiparticle energy dispersion in Eqs.~\eqref{eq:Energy_dispersion_mus_general} or ~\eqref{eq:Energy_dispersion_mus_final} does not take the form of a simple shift of a momentum-dependent energy by a chemical potential. Thus, while $\mu_\Sigma$ plays a role formally reminiscent of a chemical potential, it cannot be interpreted in the same thermodynamic sense. Instead, it should be regarded as an external parameter that controls the degree of spin polarization in the system.

Nevertheless, there exists a region of parameter space in which $\mu_\Sigma$ may
be interpreted, albeit in a rather loose sense, as playing a role analogous to
that of a chemical potential. To see this point, let us consider the energy dispersion~\eqref{eq:Energy_dispersion_mus_general} for a quasiparticle with a momentum aligned with the direction of the spin, ${\bf p} = p \, {\bf n}_\Sigma \equiv p \, {\bf e}_z$. Then, in the notations of Eq.~\eqref{eq:Energy_dispersion_mus_final}, $E_{\mathbf{p}}^{(s)} = |E_{p_z} - s \mu_\Sigma|$, implying that a relatively small spin potential, $|\mu_\Sigma| < 2E_{p_z}$, acts like a chemical potential for spin-polarized fermions propagating along the direction of the spin polarization, $E_{\mathbf{p}}^{(s)} = E_{p_z} - s \mu_\Sigma$.

Consequently, the spin potential $\mu_\Sigma$ may be interpreted, similarly to a chemical potential, as the energy cost associated with adding (or removing) a unit of spin ($\hbar$ in dimensional units) provided two conditions are satisfied: (i) the single-particle energy exceeds $|\mu_\Sigma/2|$; and (ii) the momentum of the particle is aligned (or anti-aligned, respectively) with the direction of the spin polarization so that ${\bf p} \| {\bf n}_\Sigma$. Since fermions and anti-fermions carry a half of the unit of spin, $\hbar/2$, the spin potential adds or removes the energy $\mu_\Sigma/2$.

\subsection{Thermodynamics of the spin system}
\label{Subsec:thermodynamics}

Under the path integral formalism, we have the total partition function $\mathcal{Z} = \mathcal{Z}_\mathcal{M} \mathcal{Z}_q$, where $\mathcal{Z}_\mathcal{M}=e^{-\beta \mathcal{V}_{3d} V_{\mathcal{M}}}$ is the meson contribution, with $\mathcal{V}_{3d}$ being the space volume. The quark contribution $\mathcal{Z}_q$ is given as
\begin{equation}
 \mathcal{Z}_q = \int [i d\psi^\dagger] [d\psi] e^{-S_E}, \quad
 S_E = \int\limits_0^\beta d\tau \int d^3x \mathcal{L}_E,
 \label{eq:Z}
\end{equation}
where the Euclidean Lagrangian density satisfies $\mathcal{L}_E = -\mathcal{L}_q = \psi^\dagger (\partial_\tau + H_q)\psi$.
The functional integration can be performed using the normal mode decomposition of the fermion field,
\begin{equation}
 \psi(X) = \frac{1}{\sqrt{\mathcal{V}_{3d}}} \sum_n \sum_{s,\varsigma} \sum_{\mathbf{p}} e^{i(\omega_n \tau + \mathbf{p} \cdot \mathbf{x})} \psi_{n,\varsigma}^{(s)}(\mathbf{p}) \hat{b}_{n,\varsigma}^{(s)}(\mathbf{p}),
\end{equation}
with $X_\mu = (\tau, \mathbf{x})$ denoting the Euclidean four-coordinates and
$\omega_n = (2n+1) \pi T$ being the Matsubara frequencies, ensuring the anti-periodicity of the fermion wavefunction with respect to the imaginary time $\tau = i t$. The modes $\psi_{n,\varsigma}^{(s)}(\mathbf{p})$ correspond to the Hamiltonian eigenfunctions $\psi_{\varsigma}^{(s)}(\mathbf{p})$ introduced in Eq.~\eqref{eq:H_eigen}, satisfying the orthogonality relation
\begin{equation}
 [\hat{\psi}_{n,\varsigma}^{(s)}(\mathbf{p})]^\dagger
 \hat{\psi}_{n,\varsigma'}^{(s')}(\mathbf{p}) = \delta_{ss'} \delta_{\varsigma\varsigma'}.
\end{equation}
The Fourier coefficients $\hat{b}_{n,\varsigma}^{(s)}(\mathbf{p})$ are represented as $N_f \times N_c$ Grassmann variables.

After carrying out the space-time integration, the Euclidean action $S_E$ defined in Eq.~\eqref{eq:Z}, can be expressed as
\begin{equation}
 S_E = i \beta \sum_n \sum_{s,\varsigma} \sum_{\mathbf{p}} (\omega_n - i \varsigma E^{(s)}_{\mathbf{p}}) [\hat{b}^{(s)}_{n,\varsigma}]^\dagger (\mathbf{p}) \hat{b}^{(s)}_{n,\varsigma}(\mathbf{p}),
\end{equation}
where the energy branches $\varsigma E^{(s)}_{\mathbf{p}}$ correspond to the eigenvalues of the Hamiltonian introduced in Eq.~\eqref{eq:H_eigen}.

The quark contribution to the partition function can then be obtained using the Gauss integration formula for Grassmann variables, $\int d\hat{b}^\dagger d\hat{b}\, e^{a \hat{b}^\dagger \hat{b}} = a$, yielding
\begin{equation}
 \mathcal{Z}_q = \left[\prod_{n,\mathbf{p}} \prod_{s,\varsigma} i\beta (\omega_n - i\varsigma E^{(s)}_{\mathbf{p}})\right]^{N_f N_c},
\end{equation}
where the degeneracy factor
$N_f N_c$ accounts for $N_f$ quark flavours and $N_c$ color degrees of freedom. Taking the logarithm of the above expression, we obtain
\begin{equation}
 \ln \mathcal{Z}_q = N_f N_c \sum_{n,\mathbf{p}} \sum_{s,\varsigma}  \ln [\beta(i\omega_n + \varsigma E^{(s)}_{\mathbf{p}})]\,.
\end{equation}

The total thermodynamic potential of the spin system $\mathcal{F}$ subsequently has contributions from both the meson and quark sectors. The meson contribution can be obtained in a straightforward way as $-T\ln \mathcal{Z}_{\mathcal{M}} = \mathcal{V}_{3d} V_{\mathcal{M}}$. On the other hand, the quark contribution of the free energy $\mathcal{F}_q = -T \ln \mathcal{Z}_q = \mathcal{V}_{3d} (F^{\rm ZP}_q + F_q)$ can be decomposed into its zero-point (ZP) and thermal parts:
\begin{align}
 F^{\rm ZP}_q &= -N_cN_f \sum_s \int \frac{d^3p}{(2\pi)^3} E_{\mathbf{p}}^{(s)}, \label{eq:FqZP_1}\\
 F_q &= -2N_c N_f T \sum_s \int \frac{d^3p}{(2\pi)^3} \ln\bigl(1 + e^{-\beta E_{\mathbf{p}}^{(s)}}\bigr)\,.
 \label{eq:Fqth}
\end{align}

\section{Vanishing temperature: Zero-point renormalization}\label{sec:vac_ren}

In the mean-field approximation of a homogeneous system, discussed in Sec.~\ref{Subsec:thermodynamics}, the thermodynamic potential density $F=\mathcal{F}/\mathcal{V}_{3d}$ receives a contribution from the meson potential \eqref{eq:pot_meson} with vanishing pion field, $\vec{\pi} = 0$:
\begin{align}
 V_{\mathcal{M}} \to V_\sigma = \frac{\lambda}{4}(\sigma^2-v^2)^2 - h \sigma\,.
    \label{eq:pot_sigma_tree}
\end{align}
Keeping the field $\sigma$ as a background scalar field, the potential gets a divergent one-loop contribution from the quark sector, denoted by $F_q^{\rm ZP}$ in Eq.~\eqref{eq:FqZP_1}, which we treat via self-consistent renormalization. This section outlines the procedure of renormalization in the presence of a finite spin density.

We begin with the standard LSM$_q$ model in Subsec.~\ref{sec:vac_ren:LSMq} and subsequently incorporate the axial-vector interaction to isolate and address spin-potential-dependent divergences. In Subsections~\ref{sec:vac_ren:naive} and \ref{sec:vac_ren:model2}, we introduce two schemes: a na\"ive implementation in Subsec.~\ref{sec:vac_ren:naive}, leading to a spin-independent renormalized potential; and a spin-dependent deformation in Subsec.~\ref{sec:vac_ren:model2}, featuring a free parameter $\ell$, allowing the spin to modify the vanishing temperature limit of the theory.
In Appendix~\ref{app:rg}, we discuss the renormalization scale invariance of our renormalized potential, given in Eq.~\eqref{eq:LSM_mus_model2}, in the frame of the renormalization group flow proposed in Ref.~\cite{Brandt:2025tkg}.

\subsection{Standard \texorpdfstring{LSM$_q$}{LSMq}}
\label{sec:vac_ren:LSMq}

In standard LSM$_q$, the zero-point quark contribution to the free energy in Eq.~\eqref{eq:FqZP_1} reads
\begin{align}
 F_q^{\rm ZP} &= -2 N_c N_f I_1, \qquad\
 I_1 = \int \frac{d^3p}{(2\pi)^3} E_p.
 \label{eq:I1_def}
\end{align}
Employing dimensional regularization, as explained in Appendix~\ref{app:DR}, $I_1$ evaluates to \cite{Skokov:2010sf}
\begin{equation}
 I_1 = -\frac{g^4\sigma^4}{32\pi^2}\left[\frac{1}{\epsilon}+\frac{3}{2}-\gamma_E+\ln\left(\frac{4\pi M^2}{g^2\sigma^2}\right)\right],
 \label{eq:I1_reg}
\end{equation}
where $\gamma_E$ is the Euler-Mascheroni constant and $M$ is the renormalization scale introduced in the minimal subtraction scheme.

In order to self-consistently account for this vacuum contribution coming from the fermion loop, we employ a bare meson potential of the form
\begin{equation}
 V_\sigma^{\rm bare} =
 \frac{\lambda_{0}}{4}(\sigma^2 -v_0^2)^2 -h \sigma + L^{(4)}_{0} \sigma^4 \ln \frac{f_\pi^2}{\sigma^2} + V_0.
 \label{eq:LSM_Vb}
\end{equation}
Compared to Eq.~\eqref{eq:pot_sigma_tree}, the above expression displays a logarithmic term, introduced to balance the analogous term appearing in $I_1$ above. We now demand that, at zero temperature, the renormalized potential $V^{\rm ren}_\sigma = V_\sigma^{\rm bare} + F_q^{\rm ZP}$ reduces to:\footnote{Ref.~\cite{Skokov:2010sf} argues that the $\sigma^4 \ln \sigma$ term is required to ensure that the chiral transition is a crossover in the chiral limit of the theory. With our choice of parameters, the transition is a crossover even without this term. Therefore, for simplicity, we do not keep it in the renormalized potential.}
\begin{equation}
 V_\sigma^{\rm ren} = V_\sigma(\sigma) - V_\sigma(f_\pi), \label{eq:LSM_model_def}
\end{equation}
where $V_\sigma(\sigma)$ is the classical meson form given in Eq.~\eqref{eq:pot_sigma_tree} and we subtracted the value of the potential in the vacuum (where $\sigma = f_\pi$) to ensure that $P = -(V_\sigma^{\rm ren} + F_q)$ represents the mechanical pressure.
It is straightforward to identify:
\begin{gather}
 \lambda_0 = \lambda - \frac{g^4 N_c N_f}{4\pi^2} \left(\frac{1}{\epsilon} + \frac{3}{2} -\gamma_E + \ln \frac{4\pi M^2}{g^2 f_\pi^2}\right)\,, \nonumber\\
 v_0^2 = -\frac{4\pi^2 \epsilon}{g^4 N_c N_f} \lambda v^2\,, \qquad\
 L^{(4)}_0 = -\frac{g^4 N_c N_f}{16\pi^2}\,, \nonumber\\
 V_0 =
 \frac{\lambda f_\pi^2}{4}(2v^2 - f_\pi^2) + h f_\pi\,,
 \label{eq:LSM_model}
\end{gather}
where we took into account that, while  $\lambda_0 v_0^2 = \lambda v^2$ is finite, $\lambda_0 v_0^4 = 0$, because $v_0^2 = O(\epsilon)$ vanishes when the limit $\epsilon \to 0$ is taken.

%%%%%%%%%%%%%%%%%%%%%%%%%%%%%%%%%%%%%%%%%%%%%%%%%%%%%%%%%%%%%%%%%%%%%%%%%%%%%%%%%%%%%%%%%%%%%%%%%%%%%%%%%%%%%%%%%%%%%%%%%%%%%%%%%%%%%%%%%%%%%%%%%%%%%%%%%%%%%%%%%%%%%%%%%%%%%%%%%%%%%%%%%%%%%%%%%%%%%%%%%%%%%%%%%%

\subsection{Spin-deformed \texorpdfstring{LSM$_q$}{lsmq}: a na\"ive approach}
\label{sec:vac_ren:naive}

Next, we turn our focus to the modifications due to the inclusion of the spin potential $\mu_\Sigma$.
Contrary to the case without spin, the zero-point energy $F_q^{\rm ZP}$ in Eq.~\eqref{eq:FqZP_1} exhibits an explicit dependence on the spin potential.
As pointed out in Ref.~\cite{Farias:2016let}, the arbitrariness of the regularization methods should apply only to vacuum integrals, i.e., integrals that do not depend on the parameters of the medium, $T$ and $\mu_\Sigma$.
We therefore write
\begin{multline}
 \sum_{s} E_{\mathbf{p}}^{(s)} = 2E_p + \frac{E_p^2-E_{p_z}^2}{4E_p^3}\mu_\Sigma^2 \\
 -\frac{(E_p^2 - E_{p_z}^2)^2-4E_{p_z}^2 (E_p^2 - E_{p_z}^2)}{64 E_p^7}\mu_\Sigma^4 + \mathcal{O}(\mu_\Sigma^6)
\end{multline}
and subsequently express the zero-point quark contribution to the free energy as
\begin{equation}
    F_q^{\rm ZP} = -2N_cN_f(I_1 + \mu_\Sigma^2 I_2 + \mu_\Sigma^4 I_3 + \delta I_\Sigma),
    \label{eq:IS_def}
\end{equation}
where $I_1$ is defined in Eq.~\eqref{eq:I1_def} and we introduce two more integrals, explicitly given as
\begin{subequations}
\begin{align}
 I_2 &= \int \frac{d^3p}{(2\pi)^3}  \frac{E_p^2-E_{p_z}^2}{8E_p^3}, \label{eq:I2_def}\\
 I_3 &= -\int \frac{d^3p}{(2\pi)^3}  \frac{(E_p^2-E_{p_z}^2)(E_p^2 - 5E_{p_z}^2)
 %-4E_{p_z}^2(E_{\bf p}^2-E_{p_z}^2)
 }{128 E_p^7},
 \label{eq:I3_def}
\end{align}
\end{subequations}
while $\delta I_{\Sigma}$ denotes the regular remainder. After dimensional regularization, these integrals reduce to (see Appendix~\ref{app:DR})
\begin{align}
 I_2 &= -\frac{g^2\sigma^2}{32\pi^2}\left(\frac{1}{\epsilon} - \gamma_E + \ln \frac{4\pi M^2}{g^2 \sigma^2}\right), %\label{eq:I2_reg}\\
 &
 I_3 &= \frac{1}{384\pi^2}, \label{eq:I2I3_reg}
\end{align}
while the remainder takes the form (cf.~Appendix~\ref{app:DR})
\begin{multline}
 \delta I_\Sigma = -\frac{\theta_\Sigma}{48\pi^2} \bigg[|\mu_\Sigma| p_\Sigma(\tfrac{1}{4} \mu_\Sigma^2 + \tfrac{13}{2} g^2 \sigma^2) \\
 - 3 g^2 \sigma^2(\mu_\Sigma^2 + g^2\sigma^2) \ln \frac{\frac{1}{2} |\mu_\Sigma| + p_\Sigma}{g|\sigma|}
 \bigg],
 \label{eq:dIS_result}
\end{multline}
with $p_\Sigma = \sqrt{\frac{1}{4}\mu_\Sigma^2 - g^2 \sigma^2}$ and $\theta_\Sigma = \theta(\tfrac{1}{2} |\mu_\Sigma| - g|\sigma|)$.
This step function indicates that $\delta I_\Sigma$ has consequences only at large $\mu_\Sigma$, or at low meson expectation values. Since the transition close to $\mu_\Sigma = 0$ is of crossover type, neither scenario is realized in the study of the curvature of the transition line, considered in Sec.~\ref{sec:joy:lattice}. For simplicity, this term will be subtracted from the effective potential during renormalization (see below).

% The equivalence between $\{I_2,I_3\} = \{-I_4,I_5\}$ essentially means that we can repeat the procedures of the chiral model for the axial-vector model with the simple replacement of $\mu_5^2 \to -\mu_\Sigma^2$.\\

A na\"ive regularization scheme would be to again enforce that the regularized potential agrees with $V_\sigma$ in Eq.~\eqref{eq:pot_sigma_tree} for any value of $\mu_\Sigma$ when $T= 0$. This requires us to employ medium-dependent bare parameters, which become functions of $\mu_\Sigma$.
Hence, we introduce
\begin{multline}
 V_\sigma^{\rm bare} =
 \frac{\lambda_{\rm n}}{4}(\sigma^2 -v_{\rm n}^2)^2 -h \sigma + L^{(4)}_{\rm n} \sigma^4 \ln \frac{f_\pi^2}{\sigma^2} \\ +L^{(2)}_{\rm n} \sigma^2 \ln \frac{f_\pi^2}{\sigma^2}
 - \frac{N_c N_f \theta_\Sigma}{24\pi^2} \bigg[|\mu_\Sigma| p_\Sigma(\tfrac{1}{4} \mu_\Sigma^2 + \tfrac{13}{2} g^2 \sigma^2) \\
 - 3 g^2 \sigma^2(\mu_\Sigma^2 + g^2\sigma^2) \ln \frac{\frac{1}{2}|\mu_\Sigma| + p_\Sigma}{g f_\pi}
 \bigg] + V_{\rm n}.
 \label{eq:LSM_mus_Vbare1}
\end{multline}
Compared to Eq.~\eqref{eq:LSM_Vb}, the above expression has an extra logarithmic term, to compensate the analogous term coming from $I_2$ in Eq.~\eqref{eq:I2I3_reg}, as well as a term involving the step function $\theta_\Sigma = \theta(\tfrac{1}{2}|\mu_\Sigma| - g|\sigma|)$, ensuring the cancellation of the part that depends on $p_\Sigma$ in Eq.~\eqref{eq:dIS_result}. Imposing Eq.~\eqref{eq:LSM_model_def} again, we arrive at
\begin{gather}
 v_{\rm n}^2 = v_0^2 - \frac{\mu_\Sigma^2}{4g^2}(2 - 3\epsilon) - \frac{2\pi^2 \lambda \mu_\Sigma^2 \epsilon}{g^6 N_c N_f}\,, \nonumber\\
 L^{(2)}_{\rm n} = -\frac{N_c N_f g^2 \mu_\Sigma^2}
 {16\pi^2} \tilde{\theta}_\Sigma, \qquad\
 L^{(4)}_{\rm n} = -\frac{g^4 N_c N_f}
 {16\pi^2} \tilde{\theta}_\Sigma\,, \nonumber\\
 V_{\rm n} = V_0 -
 \frac{\lambda_{\rm n} v_{\rm n}^4}{4} + \frac{N_c N_f \mu_\Sigma^4}{192\pi^2}\,,
 \label{eq:LSM_mus_model1}
\end{gather}
while $\lambda_{\rm n} = \lambda_0$ from Eq.~\eqref{eq:LSM_model} and the step function $\tilde{\theta}_\Sigma = 1 - \theta_\Sigma$ is non-vanishing only when $g|\sigma| > \tfrac{1}{2}|\mu_\Sigma|$.
In contrast to Eq.~\eqref{eq:LSM_model}, the term $\lambda_{\rm n} v_{\rm n}^4$ no longer vanishes, as $v_{\rm n}^2 = - \mu_\Sigma^2 / 2g^2 + O(\epsilon)$. Hence, the offset $V_{\rm n}$ diverges as $\epsilon \to 0$.
In the scheme discussed above, the bare parameters of the meson potential exhibit an explicit medium dependence, such that the potential after absorbing the zero-point terms becomes medium-independent. However, as we will show in Sec.~\ref{sec:res}, this regularization scheme leads to results which are inconsistent with lattice data. We therefore consider an extension of this procedure, described below.

\subsection{Renormalization with spin effects}\label{sec:vac_ren:model2}

It is conceivable that, after renormalization, the meson potential acquires a dependence on $\mu_\Sigma$. We retain the restriction that at $\mu_\Sigma = 0$, Eq.~\eqref{eq:LSM_model_def} is satisfied.
Furthermore, at vanishing temperature, we impose that the point $\sigma = f_\pi$ corresponds to a (local) minimum and that the $\sigma$ meson mass remains unchanged:
\begin{equation}
 \left.\frac{\partial F}{\partial \sigma}\right|_{\substack{T = 0 \\
 \sigma = f_\pi}} = 0, \quad
 \left.\frac{\partial^2 F}{\partial \sigma^2}\right|_{\substack{T = 0 \\
 \sigma = f_\pi}} = m_\sigma^2,
 \label{eq:scheme2}
\end{equation}
where $m_\sigma = \sqrt{\lambda(3f_\pi^2 - v^2)} = 600$
MeV (the vacuum mass of the $\sigma$ meson).
To satisfy the above equations, we write the bare potential as
\begin{multline}
 V_\sigma^{\rm bare} =
 \frac{\lambda_{\rm s}}{4}(\sigma^2 -v_{\rm s}^2)^2 -h \sigma + L^{(4)}_{\rm s} \sigma^4 \ln \frac{f_\pi^2}{\sigma^2} \\ +L^{(2)}_{\rm s} \sigma^2 \ln \frac{f_\pi^2}{\sigma^2}
 - \frac{N_c N_f \theta_\Sigma}{24\pi^2} \bigg[|\mu_\Sigma| p_\Sigma(\tfrac{1}{4} \mu_\Sigma^2 + \tfrac{13}{2} g^2 \sigma^2) \\
 - 3 g^2 \sigma^2(\mu_\Sigma^2 + g^2\sigma^2) \ln \frac{\frac{1}{2}|\mu_\Sigma| + p_\Sigma}{g f_\pi}
 \bigg] + V_{\rm s}.
 \label{eq:LSM_mus_Vbare2}
\end{multline}
In this second renormalization prescription, we consider that the parameters with the subscript ``s'' represent a $\mu_\Sigma$-dependent deformation of those in Eq.~\eqref{eq:LSM_mus_Vbare1}.  For simplicity, we control this deformation by allowing the coefficient $L^{(2)}_{\rm s} = L_{\rm n}^{(2)} + \Delta L_{\rm s}^{(2)}$ to deviate from $L_{\rm n}^{(2)}$ as
\begin{equation}
 \Delta L^{(2)}_{\rm s} = \frac{N_c N_f g^2 \mu_\Sigma^2}{16\pi^2} \ell\,,
 \label{eq:L2scheme2}
\end{equation}
where $\ell$ is a free parameter. Note that, contrary to $L^{(2)}_{\rm n}$, $\Delta L^{(2)}_{\rm s}$ is non-vanishing regardless of the relation between $\mu_\Sigma$ and $g \sigma$.
The parameters $(\lambda_{\rm s}, v_{\rm s}^2, V_{\rm s})$ in Eq.~\eqref{eq:LSM_mus_Vbare2} receive $\ell$-dependent corrections with respect to $(\lambda_{\rm n}, v_{\rm n}^2, V_{\rm n})$ in Eq.~\eqref{eq:LSM_mus_model1}, denoted $(\Delta \lambda_{\rm s}, \Delta v_{\rm s}^2, \Delta V_{\rm s})$, i.e.
\begin{equation}
 \lambda_{\rm s} = \lambda_{\rm n} + \Delta \lambda_{\rm s}, \quad
 v^2_{\rm s} = v_{\rm n}^2 + \Delta v^2_{\rm s}, \quad
 V_{\rm s} =V_{\rm n} + \Delta V_{\rm s}.
 \label{eq:parascheme2}
\end{equation}

Imposing Eqs.~\eqref{eq:scheme2} gives
\begin{gather}
 \Delta \lambda_{\rm s} = \frac{N_c N_f g^2 \mu_\Sigma^2}{16\pi^2 f_\pi^2}\ell, \quad
 \Delta v_{\rm s}^2 = -\frac{\mu_\Sigma^4 \ell \epsilon}{4 g^4 f_\pi^2}, \nonumber\\
 \Delta V_{\rm s} = - \frac{\ell N_c N_f \mu_\Sigma^2}{32\pi^2 g^2 f_\pi^2} (g^4 f_\pi^4 - \tfrac{1}{4}\mu_\Sigma^4).
\end{gather}
In our proposed scheme, $L^{(4)}_{\rm s} = L^{(4)}_{\rm n}$.
The resulting renormalized meson potential $V_\sigma^{\rm ren} = V_\sigma^{\rm bare} + F_q^{\rm ZP}$ reads:
\begin{multline}
 V_\sigma^{\rm ren} = \frac{\lambda}{4}[(\sigma^2 - v^2)^2 - (f_\pi^2 - v^2)^2] - h (\sigma - f_\pi) \\
 + \frac{\ell N_c N_f g^2 \mu_\Sigma^2}{32 \pi^2 f_\pi^2} \left(\sigma^4 - f_\pi^4 + 2f_\pi^2 \sigma^2 \ln \frac{f_\pi^2}{\sigma^2}\right).
 \label{eq:LSM_mus_model2}
\end{multline}

An important consequence of
Eq.~\eqref{eq:LSM_mus_model2} is that the coefficient of the $\sigma^4$ term in the meson potential becomes $\mu_\Sigma$-dependent. Crucially, the thermodynamic stability of the system requires that this coefficient be positive. This translates into the constraint $\ell > 0$. If, on the contrary, $\ell < 0$, then the sign of the coefficient of $\sigma^4$ becomes negative when the spin potential $\mu_\Sigma$ exceeds the following threshold:
\begin{equation}
  \mu_\Sigma > \left(-\frac{8 \pi^2 \lambda f_\pi^2}{g^2 N_c N_f \ell}\right)^{1/2} \qquad {\rm for} \quad
  \ell < 0\,.
\end{equation}

The na\"ive scheme discussed in Subsec.~\ref{sec:vac_ren:naive} is recovered when $\ell = 0$. We view $\ell > 0$ as a free parameter of the scheme, whose value we will employ in Subsec.~\ref{sec:joy:lattice} to fit the lattice data reported in Ref.~\cite{Braguta:2025ddq}.

%%%%%%%%%%%%%%%%%%%%%%%%%%%%%%%%%%%%%%%%%%%%%%%%%%%%%%%%%%%%%%%%%%%%%%%%%%%%%%%%%%%%%%%%%%%%%%%%%%%%%%%%%%%%%%%%%%%%%%%%%%%%%%%%%%%%%%%%%%%%%%%

\subsection{Phase structure}
\label{sec:vac_ren:phase}

\begin{figure}
%\begin{figure}[!htb]
\centering
\begin{tabular}{c}
\includegraphics[width=.95\columnwidth]{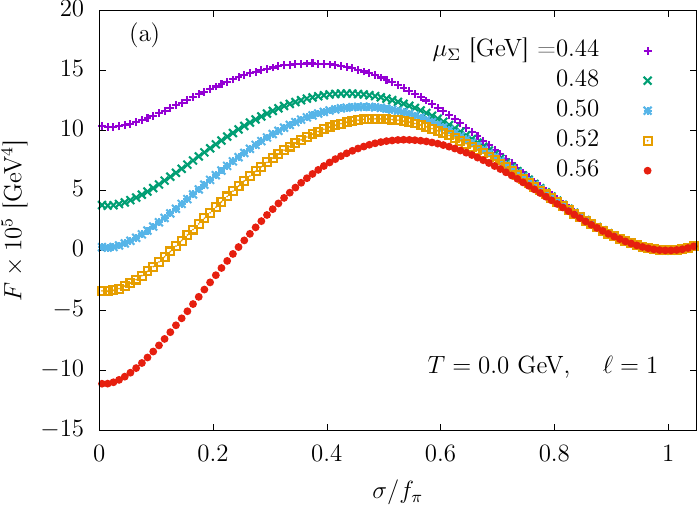} \\
 \includegraphics[width=0.95\columnwidth]{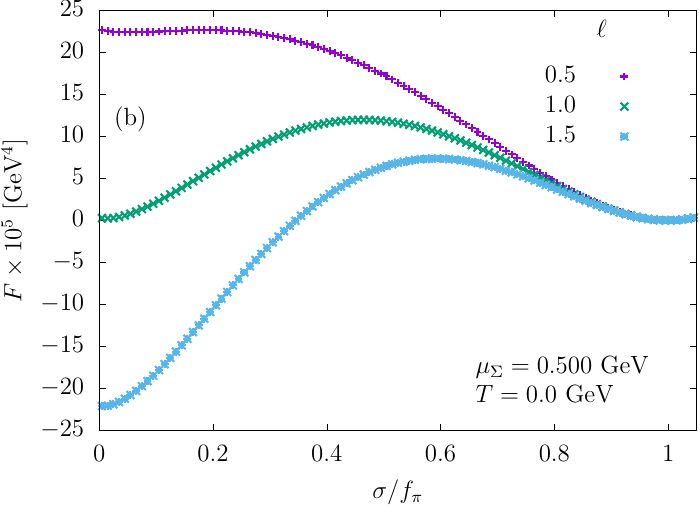}
\end{tabular}
\caption{The dependence of the vanishing-temperature thermodynamic potential density $F =V_\sigma^{\rm ren}$ \eqref{eq:LSM_mus_model2} on $\sigma$ for (a) various values of the spin potential $\mu_\Sigma$ at fixed $\ell = 1$; and (b) various values of $\ell$ at fixed $\mu_\Sigma=0.5$~GeV.
}
\label{fig:F_vac}
\end{figure}

In the mean-field approximation, the expectation value of the $\sigma$ meson is obtained by demanding the minimization of the free energy. Figure~\ref{fig:F_vac} shows the thermodynamic potential density at vanishing temperature, $F = V^{\rm ren}_\sigma$. In panel (a), we fixed $\ell = 1$ and varied the spin potential, $\mu_\Sigma$. At low values of $\mu_\Sigma$, the dominant minimum of $V^{\rm ren}_\sigma$ is at $\sigma = f_\pi$ and the system is in the chirally-broken state. As $\mu_\Sigma$ increases, a second minimum of $V^{\rm ren}_\sigma$ close to $\sigma \simeq 0$ becomes dominant and chiral symmetry is restored. Panel (b) shows that, at fixed $\mu_\Sigma = 0.5$~GeV, increasing $\ell$ promotes chiral restoration, leading to the appearance of the second minimum, which eventually becomes dominant.

\begin{figure}
%\begin{figure}[!htb]
\centering
\begin{tabular}{c}
\includegraphics[width=0.95\columnwidth]{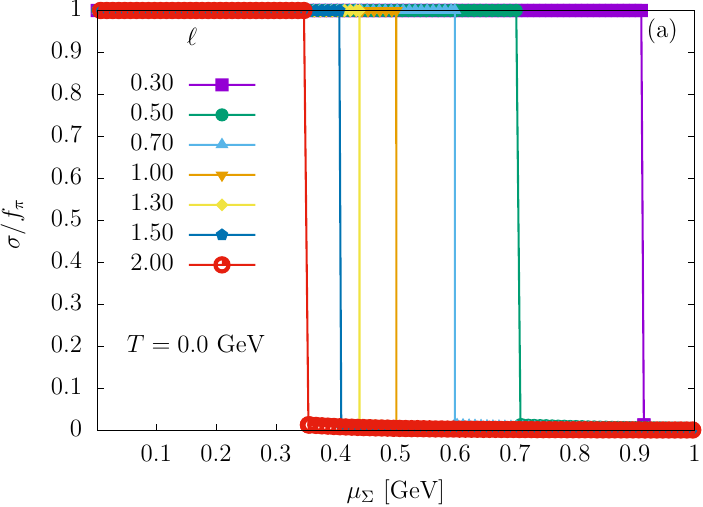} \\
 \includegraphics[width=0.95\columnwidth]{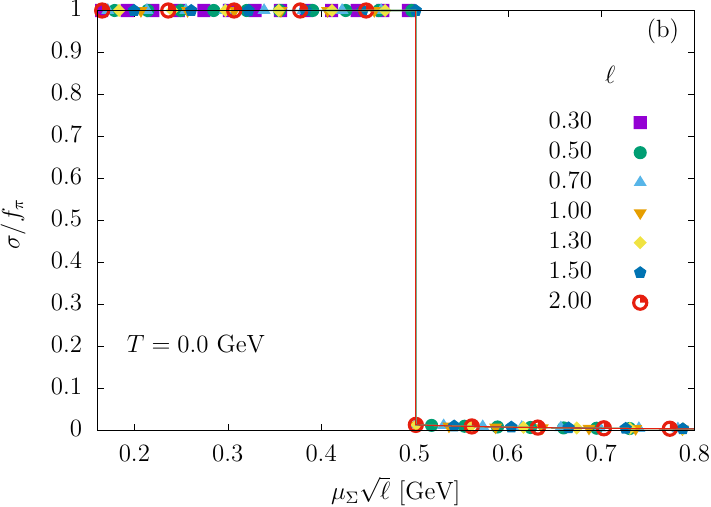}
\end{tabular}
\caption{The dependence of $\sigma/f_\pi$ on (a) $\mu_\Sigma$
and (b) $\mu_\Sigma\sqrt{\ell}$, for different values of $\ell$ and at $T=0$.
}
\label{fig:sigma_vac}
\end{figure}

To pinpoint the value of $\sigma$ that minimizes the thermodynamic potential density at $T=0$, we impose the saddle point equation
\begin{multline}
 \frac{\partial V_\sigma^{\rm ren}}{\partial \sigma} = \lambda(\sigma^2 - v^2)\sigma - h \\
 + \frac{\ell N_c N_f \mu_\Sigma^2 g^2 \sigma}{8\pi^2 f_\pi^2}\left(\sigma^2 + f_\pi^2 \ln \frac{f_\pi^2}{\sigma^2} - f_\pi^2\right) = 0.
\end{multline}
Figure~\ref{fig:sigma_vac} shows the variation of $\sigma/f_\pi$ with respect to the spin potential $\mu_\Sigma$. For $\mu_\Sigma < \mu_\Sigma^c$, the system remains in the chirally-broken phase. When $\mu_\Sigma > \mu_\Sigma^c$, a first-order phase transition to the chirally-restored phase can be observed. Since at $T = 0$, $\partial V^{\rm ren}_\sigma / \partial \sigma$ depends on $\ell$ and $\mu_\Sigma$ only through the combination $\ell \mu_\Sigma^2$, the critical spin potential obeys the scaling $\mu_\Sigma^c \sqrt{\ell} = {\rm const}$. We test this hypothesis in panel (b) of Fig.~\ref{fig:sigma_vac} and we observe that this constant equals $\mu_\Sigma \sqrt{\ell} \simeq 0.5$ GeV.

We can get a better understanding of the chiral restoration transition by considering a small-$\sigma$ expansion of the renormalized meson potential $V_\sigma^{\rm ren}$ in Eq.~\eqref{eq:LSM_mus_model2},
\begin{equation}
 V_\sigma^{\rm ren} = c - h \sigma -a \sigma^2 - b \sigma^2 \ln \frac{\sigma}{f_\pi} + O(\sigma^3),
\end{equation}
where
\begin{gather}
 c = \frac{\lambda f_\pi^2}{4} (2v^2 -f_\pi^2) + h f_\pi - \frac{\ell N_c N_f \mu_\Sigma^2 g^2 f_\pi^2}{32\pi^2}, \nonumber\\
 a = \frac{\lambda}{2} v^2, \quad
 b = \frac{\ell N_c N_f \mu_\Sigma^2 g^2}{8\pi^2}.
\end{gather}
The saddle point equation reduces to
\begin{equation}
 \frac{\partial V_\sigma^{\rm ren}}{\partial \sigma} = -h -(2a + b) \sigma - 2b \sigma \ln \frac{\sigma}{f_\pi} = 0,
\end{equation}
from where we can deduce that $\sigma \ln(\sigma / f_\pi) = -[h + (2a + b)\sigma] / (2b)$. At the minimum of the thermodynamic potential, we have
\begin{equation}
 V^{\rm ren}_\sigma = c - \frac{h \sigma}{2} + \frac{b \sigma^2}{2}.
\end{equation}
Demanding that the phase transition happens at the above point is equivalent to imposing $V^{\rm ren}_\sigma = 0$, which leads to
\begin{equation}
 \sigma = \frac{h - \sqrt{h^2 - 8bc}}{2b}.
\end{equation}
Imposing now that $\sigma \ln (\sigma / f_\pi) = -[h + (2a + b)\sigma] / (2b)$ yields the solution $\mu_\Sigma \sqrt{\ell} \simeq 0.50$ GeV, in agreement with panel (b) of Fig.~\ref{fig:sigma_vac}.

\subsection{Pressure and spin density}\label{sec:vac_ren:observables}

\begin{figure}
\begin{center}
\includegraphics[width=\linewidth]{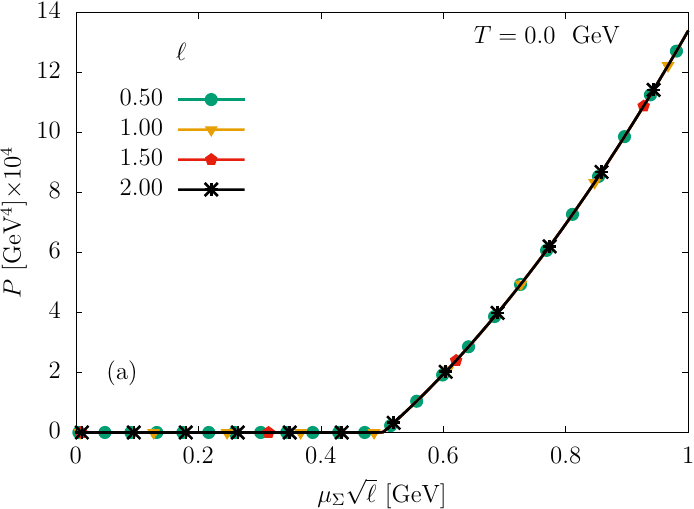}\\
\includegraphics[width=\linewidth]{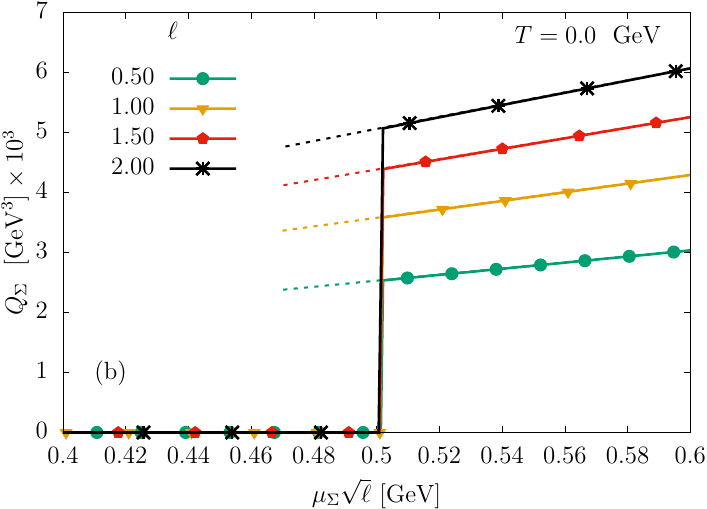}
\end{center}
\caption{
(a) Pressure and (b) spin density with respect to $\mu_\Sigma \sqrt{\ell}$, at vanishing temperature $T = 0$, for various values of $\ell$.
The dashed lines represent the asymptotic form of the spin density in the chirally-restored phase, as given in Eq.~\eqref{eq:QZP_as}.
}
\label{fig:T0_spin}
\end{figure}

We now discuss the system pressure, $P = -F= -V^{\rm ren}_\sigma - F_q$. At vanishing temperature, the thermal quark contribution vanishes, $F_q = 0$, and the pressure receives only the zero-point contribution, $P = P^{\rm ZP} = -V^{\rm ren}_\sigma$. In the chirally-broken state, $\sigma = f_\pi$ and $V^{\rm ren}_\sigma$ vanishes, see Eq.~\eqref{eq:LSM_mus_model2}, therefore $P^
{\rm ZP}(\sigma = f_\pi) = 0$. As discussed in Sec.~\ref{sec:vac_ren:phase}, the chiral restoration takes place when $\mu_\Sigma \sqrt{\ell} \simeq 0.5$ GeV. By the Gibbs construction, at this point the pressure must be equal to its vacuum value, $P = 0$, increasing as $\mu_\Sigma^2 \ell$. Since in the chirally-restored phase, $\sigma \simeq 0$, the scaling with $\mu_\Sigma \sqrt{\ell}$ of the chiral restoration transition transfers also to the pressure, as can be seen in Fig.~\ref{fig:T0_spin}(a).

At finite spin potential $\mu_\Sigma$, the system can develop a non-vanishing spin density $Q_\Sigma = \langle \mathcal{S}^z \rangle$ corresponding to the spin density operator $\mathcal{S}^z = \psi^{\dagger} \Sigma^z \psi$ along the $z$ direction. In the grand canonical ensemble, the spin density can be obtained via
\begin{equation}
 Q_\Sigma = -\frac{\partial F}{\partial \mu_\Sigma}\,.
\end{equation}
In general, this quantity can be represented as the sum of two terms, $Q_\Sigma = Q_\Sigma^{\rm ZP} + Q_\Sigma^q$, corresponding to the contributions from the renormalized meson potential at zero temperature, $Q_\Sigma^{\rm ZP} = -\partial V_\sigma^{\rm ren} / \partial \mu_\Sigma$, and from the quark thermal potential, $Q_\Sigma^q = -\partial F_q / \partial \mu_\Sigma$, respectively. Focusing on the contribution at vanishing temperature, when $F = V_\sigma^{\rm ren}$ and
\begin{equation}
 Q^{\rm ZP}_\Sigma = \frac{\ell N_c N_f g^2 \mu_\Sigma}{16\pi^2f_\pi^2} \left(f_\pi^4 - 2 f_\pi^2\sigma^2\ln \frac{f_\pi^2}{\sigma^2} - \sigma^4\right),
 \label{eq:QZP}
\end{equation}
it can be seen that $Q^{\rm ZP}_\Sigma$ vanishes in the chirally-broken phase, when $\sigma = f_\pi$. In the chirally-restored phase, the condensate is very small, $\sigma \ll f_\pi$, and Eq.~\eqref{eq:QZP} implies that $Q_\Sigma^{\rm ZP}$ increases linearly with the spin potential:
\begin{equation}
 Q^{\rm ZP}_\Sigma \simeq \frac{\ell N_c N_f}{16\pi^2} g^2 f_\pi^2 \mu_\Sigma.
 \label{eq:QZP_as}
\end{equation}
Contrary to the pressure and the point of the first-order phase transition, which scale as $\mu_\Sigma \sqrt{\ell}$, the spin density $Q_\Sigma$ scales as $\mu_\Sigma \ell$. Figure~\ref{fig:T0_spin}(b) shows $Q_\Sigma$ as a function of $\mu_\Sigma \sqrt{\ell}$, for various values of $\ell$.

\section{Finite temperature}
\label{sec:joy}

After renormalizing the zero-point contribution to the free energy, in this section we turn our attention to its thermal part and examine certain subtle aspects of the energy dispersion relation in the presence of a finite spin potential.

\subsection{Thermal contribution to the thermodynamic potential}\label{sec:joy:Fq}

We start by writing the thermal part $F_q$ of the quark free energy, given in Eq.~\eqref{eq:Fqth}, with respect to spherical coordinates:
\begin{multline}
 F_q = -\frac{N_c N_f T}{2\pi^2} \sum_s \int\limits_0^\infty dp\, p^2 \int\limits_{-1}^1 d \cos\theta \\\times
 \ln(1 + e^{-\beta E_{\mathbf{p}}^{(s)}}).
 \label{eq:Fqth_sph}
\end{multline}
In order to study the thermodynamic properties of the system, the mean value of the $\sigma$ meson is obtained by enforcing the saddle point equation:
\begin{equation}
 \frac{\partial F}{\partial \sigma} = \frac{\partial V_\sigma^{\mathrm{ren}}}{\partial \sigma} + \frac{\partial F_q}{\partial \sigma} = 0,
\end{equation}
where $V_\sigma^{\mathrm{ren}}$ is introduced in Eq.~\eqref{eq:LSM_mus_model2}.
We now write down the expression of $\partial F_q /\partial \sigma$ with respect to the spherical coordinates introduced above:
\begin{multline}
    \frac{\partial F_q}{\partial \sigma}=\frac{N_cN_fg^2 \sigma}{2\pi^2}  \sum_s \int\limits_0^\infty dp\, p^2 \int\limits_{-1}^1 \frac{d \cos\theta}{E_{\mathbf{p}}^{(s)}(e^{\beta E_{\mathbf{p}}^{(s)}} + 1)}\\\times
    \left(1 - \frac{s \mu_\Sigma}{E_{p_z}}\right).
    \label{eq:dFds_sph}
\end{multline}

The presence of $E_{\mathbf{p}}^{(s)}$ in the denominator of Eq.~\eqref{eq:dFds_sph} reveals an unexpected infrared divergence. As can be seen from Eq.~\eqref{eq:Energy_dispersion_mus_final}, $E_{\mathbf{p}}^{(s)}$ vanishes when $s \mu_\Sigma = E_{p_z} = \sqrt{p^2 \cos^2\theta + g^2 \sigma^2}$ and $p_\perp = 0$. This integrable singularity can be removed by changing the integration variable from $p$ to $E_{\mathbf{p}}^{(s)} \to E$, as discussed in Appendix~\ref{app:sad}. Without repeating the details here, several mathematical manipulations allow Eqs.~\eqref{eq:Fqth_sph} and \eqref{eq:dFds_sph} to be put in the following compact form:
\begin{subequations}
\label{eq:suitcaseless}
\begin{align}
 F_q &= -\frac{N_c N_f T}{\pi^2} \sum_{s} \int\limits_{g\sigma}^\infty dE\, \widetilde{E}_s p \ln (1 + e^{-\beta |\widetilde{E}_s|}), \label{eq:F} \\
 \frac{\partial F_q}{\partial \sigma} &= \frac{N_c N_f}{\pi^2} g^2 \sigma \sum_{s} \int\limits_{g\sigma}^\infty \frac{dE}{e^{\beta |\widetilde{E}_s|} + 1} \nonumber\\
 & \qquad\qquad \times {\rm sgn}(\widetilde{E}_s) \left[p - s \mu_\Sigma \ln \left(\frac{E + p}{g\sigma}\right)\right], \label{eq:dFds}
\end{align}
\end{subequations}
where $s = \pm 1/2$, $p = \sqrt{E^2 - g^2\sigma^2}$ and $\widetilde{E}_s = E - s\mu_\Sigma$.

%%%%%%%%%%%%%%%%%%%%%%%%%%%%%%%%%%%%%%%%%%%%%%%%%%%%%%%%%%%%%%%%%%%%%%%%%%%%%%%%%%%%%%%%%%%%%%%%%%%%%%%%%%%%%%%%%%%%%%%%%%%%%%%%%%%%%%%%%%%%%%%%

The modulus $|\widetilde{E}_s|$ appearing in Eq.~\eqref{eq:F} indicates that the spin potential $\mu_\Sigma$ does not introduce a Fermi level. It is easy to check that $\lim_{\beta \to \infty} F_q = 0$, regardless of the value of $\mu_\Sigma$. To gain more insight into the properties of $F_q$, we employ integration by parts to arrive at
\begin{multline}
 F_q = -\frac{N_c N_f}{\pi^2} \sum_{s} \int_{g\sigma}^\infty \frac{dE\, {\rm sgn}(\widetilde{E}_s)}{e^{\beta|\widetilde{E}_s|} + 1} \left(\frac{p^3}{3} - \frac{pE}{2} s \mu_\Sigma \right. \\
 \left. + \frac{1}{2} s\mu_\Sigma g^2 \sigma^2 \ln \left(\frac{E + p}{g\sigma}\right)\right).
 \label{eq:F_by_parts}
\end{multline}
Keeping in mind that $\widetilde{E}_s = E - s\mu_\Sigma$, the following relation,
\begin{equation}
 \frac{{\rm sgn}(\widetilde{E}_s)}{e^{\beta|\widetilde{E}_s|} + 1} = \frac{1}{e^{\beta \widetilde{E}_s} + 1} - \theta(-\widetilde{E}_s),
 \label{eq:FD_wsignEs_relation}
\end{equation}
shows that $\mu_\Sigma$ acts as the usual baryon (vector) chemical potential, while the step function serves to subtract the $T\to 0$ contribution, thereby eliminating the Fermi level contributions from the corresponding quantities. The first two terms in Eq.~\eqref{eq:F_by_parts} are easily recognizable as the pressure $P_{\rm FD}(\mu, T)$ and charge density $Q_{\rm FD} = \partial P_{\rm FD}(\mu, T) / \partial \mu$ of a Fermi gas with temperature $T$ and usual (vector) chemical potential equal to $\mu = \mu_\Sigma/2$, allowing $F_q$ to be expressed as
\begin{equation}
 \frac{F_q}{N_c N_f} = -\Delta P_{\rm FD}(\mu, T) + \frac{1}{2} \mu \Delta Q_{\rm FD}(\mu, T) + \delta F_q,
 \label{eq:Fq_extreme}
\end{equation}
where $\Delta P_{\rm FD}(\mu, T) = P_{\rm FD}(\mu, T) - P_{\rm FD}(0,\mu)$ refers to the Fermi-Dirac pressure with the Fermi-level contribution (corresponding to $T = 0$) subtracted, and similar for $\Delta Q_{\rm FD}(\mu, T)$, while
\begin{equation}
 \delta F_q = -\frac{\mu_\Sigma T g^2 \sigma^2}{2\pi^2} \sum_{s = \pm \frac{1}{2}} s \int_{g\sigma}^\infty \frac{dE}{p} \ln(1 + e^{-\beta|\widetilde{E}_s|}).
 \label{eq:dFq_def}
\end{equation}
The term $\delta F_q$ plays a role also in the evaluation of the fermion condensate $\langle\bar\psi\psi\rangle$, defined via $\partial F_q / \partial \sigma = g\, \langle\bar\psi\psi\rangle$:
\begin{equation}
 \frac{\partial F_q}{\partial \sigma} = g N_c N_f \Delta \langle\bar\psi\psi\rangle_{\rm FD} + \frac{2}{\sigma} \delta F_q,
\end{equation}
with $\Delta \langle \bar{\psi}\psi \rangle_{\rm FD} = -g^{-1} \partial \Delta P_{\rm FD} /\partial\sigma$.

We now discuss the small-mass limit of $F_q$. The Fermi-Dirac pressure is known \cite{kapusta_gale_2006}:
\begin{multline}
 P_{\rm FD}(\mu, T) \simeq \frac{7\pi^2T^4}{180} + \frac{\mu^2 T^2}{6} + \frac{\mu^4}{12\pi^2} - \frac{g^2 \sigma^2}{12}\left(T^2 + \frac{3\mu^2}{\pi^2}\right) \\
 + \frac{g^4 \sigma^4}{8\pi^2} \left(\frac{3}{4} - \gamma_E - \ln \frac{g\sigma}{\pi T}\right),
 \label{eq:PFD_highT}
\end{multline}
where terms of combined sixth order in $g\sigma$ and $\mu_\Sigma$ were neglected. Subtracting the $T \to 0$ part of $P_{\rm FD}$ in the same limit,
\begin{equation}
 P_{\rm FD}(\mu, 0) = \frac{\mu^4}{12\pi^2} - \frac{g^2 \sigma^2 \mu^2}{4\pi^2} + \frac{g^4 \sigma^4}{8\pi^2} \left(\frac{3}{4} - \ln \frac{g\sigma}{2|\mu|}\right),
\end{equation}
we arrive at
\begin{multline}
 \Delta P_{\rm FD}(\mu, T) \simeq \frac{7\pi^2T^4}{180} + \frac{\mu^2 T^2}{6} - \frac{g^2 \sigma^2 T^2}{12} \\
 - \frac{g^4 \sigma^4}{8\pi^2} \left(\gamma_E - \ln \frac{2|\mu|}{\pi T}\right),
 \label{eq:smallm_P}
\end{multline}
where we assumed the hierarchy $g \sigma < |\mu|$, valid at large temperatures, where $g \sigma = O(T^{-2})$.
Differentiating the pressure~\eqref{eq:smallm_P} with respect to $\mu$, we get
\begin{equation}
 \Delta Q_{\rm FD}(\mu, T) \simeq \frac{\mu T^2}{3}.
 \label{eq:smallm_Q}
\end{equation}

Finally, for the term $\delta F_q$ at small $g\sigma$, we have, to leading order in $\mu_\Sigma / T$ (see Appendix~\ref{app:deltaFq}):
\begin{equation}
 \delta F_q \simeq -\frac{\mu_\Sigma^2 g^2 \sigma^2}{8\pi^2} \left(1 - \gamma_E- \ln\frac{|\mu_\Sigma|}{\pi T}\right).
 \label{eq:smallm_dFq}
\end{equation}

Substituting the small-mass expansions from Eqs.~\eqref{eq:smallm_P}, \eqref{eq:smallm_Q} and \eqref{eq:smallm_dFq} into Eq.~\eqref{eq:Fq_extreme} gives
\begin{multline}
 F_q \simeq -N_c N_f \left[\frac{7\pi^2 T^4}{180} - \frac{g^2 \sigma^2 T^2}{12} \right. \\
 + \frac{\mu_\Sigma^2 g^2 \sigma^2}{8\pi^2} \left(1 - \gamma_E- \ln\frac{|\mu_\Sigma|}{\pi T}\right)\\
 \left. - \frac{g^4\sigma^4}{8\pi^2} \left(\gamma_E- \ln \frac{|\mu_\Sigma|}{\pi T}\right) \right].
 \label{eq_Fq_mass0}
\end{multline}
It is worth remarking that, when the mass is negligible, the quark contribution to the thermodynamic potential becomes independent of $\mu_\Sigma$.

\subsection{Phase structure}
%\label{sec:sad_insights:PT}
\label{sec:joy:PT}

\begin{figure}
\includegraphics[width=\linewidth]{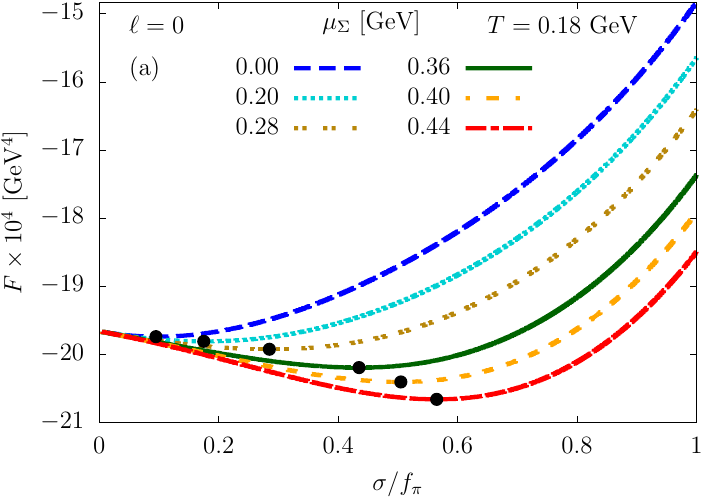} \\
\includegraphics[width=\linewidth]{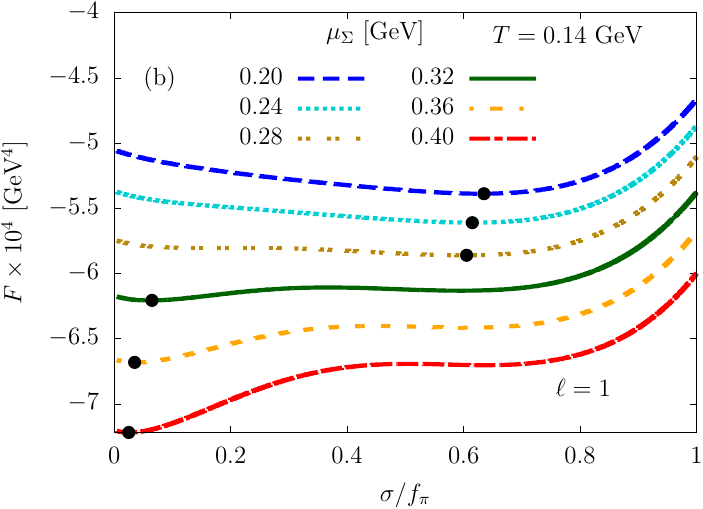}
% \end{tabular}
\caption{Thermodynamic potential $F$ as a function of $\sigma / f_\pi$, for various values of the spin potential $\mu_\Sigma$, for (a) $(\ell, T) = (0, 0.18\ {\rm GeV})$; and (b) $(\ell, T) = (1, 0.14\ {\rm GeV})$. The black blobs represent the values of $\sigma / f_\pi$ corresponding to the absolute minima of ~$F$.
}
\label{fig_F_TPT}
%\end{figure}
\end{figure}

We now consider the chiral restoration phase transition at finite temperature and spin potential. Since the renormalized free energy depends on the parameter $\ell$, in this subsection we investigate the phase structure of the model as a function of $\ell$. These results will then allow us to fix the parameter $\ell$ in the following section using the lattice data reported in Ref.~\cite{Braguta:2025ddq}.

At a vanishing spin potential, the system undergoes a crossover transition to the chirally-restored phase at the pseudocritical temperature $T_{\rm pc} = 0.147$ GeV. It is worth noticing that this temperature does not depend on the renormalization parameter $\ell$, which is associated with a finite spin potential $\mu_\Sigma$.

In Fig.~\ref{fig_F_TPT}, we show the thermodynamic potential as a function of $\sigma / f_\pi$ for various values of $\mu_\Sigma$ with two specified combinations of $\ell$ and $T$. In panel (a), we study the effect of the spin potential in the case $\ell = 0$, corresponding to the na\"ive renormalization approach discussed in Sec.~\ref{sec:vac_ren:naive}, with the temperature set at $T = 0.18$ GeV (above the crossover transition temperature). The absolute minimum of $F$ (denoted by the black blobs) shifts to higher values of $\sigma$ as $\mu_\Sigma$ is increased, signaling the breaking of chiral symmetry. In panel (b), we consider a larger value $\ell = 1$ and fix $T = 0.14$ GeV (below the pseudocritical temperature). In this case, increasing $\mu_\Sigma$ causes the minimum of $F$ to shift towards lower values of $\sigma$, indicating chiral restoration. For the particular case shown in panel (b), the chiral transition is of first order.

%%%%%%%%%%%%%%%%%%%%%%%%%%%
\begin{figure}
\includegraphics[width=\linewidth]{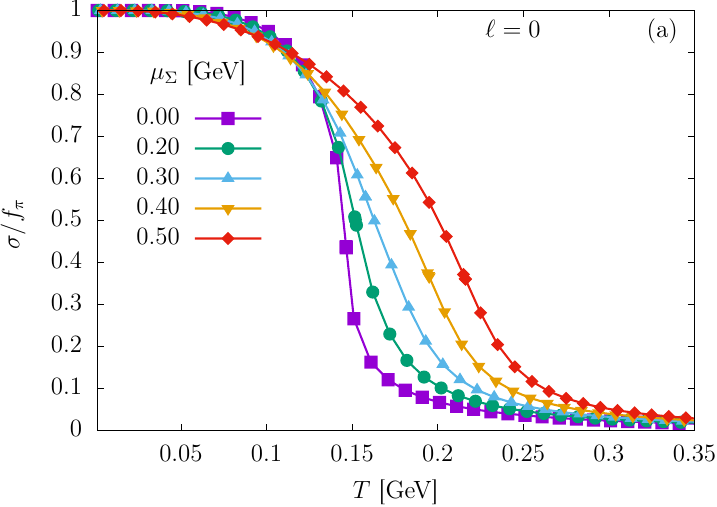} \\
\includegraphics[width=\linewidth]{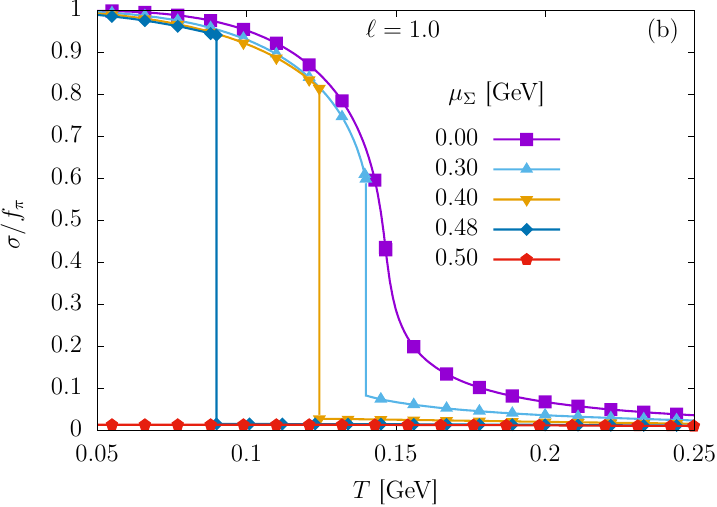}
\caption{$\sigma /f_\pi$ as a function of $T$, for various values of $\mu_\Sigma$, for (a) $\ell=0$; and (b) $\ell=1$.
}
\label{fig_PT_comp}
\end{figure}
%%%%%%%%%%%%%%%%%%%%

In what follows, we always set the mean-field value of the $\sigma$ meson corresponding to the minimum of the total thermodynamic potential. In Fig.~\ref{fig_PT_comp}, we show $\sigma/f_\pi$ as a function of $T$ for various values of $\mu_\Sigma$, with two fixed values of $\ell$. For the case $\ell = 0$, shown in panel (a), the renormalized meson potential $V_\sigma^{\rm ren}$ is independent of $\mu_\Sigma$, such that $F$ receives $\mu_\Sigma$ contributions only from the thermal quarks via $F_q$. It can be seen that increasing $\mu_\Sigma$ causes the transition point to migrate towards larger temperatures, inhibiting chiral restoration. In contrast, for the case $\ell = 1$, shown in panel (b), the thermodynamic potential receives $\mu_\Sigma$ contributions also via $V_\sigma^{\rm ren}$. We observe two effects: first, the transition point shifts to lower temperatures, indicating that the spin potential promotes chiral restoration; second, the transition becomes sharper, eventually turning into a first-order transition. Comparing panel (a) with panel (b), we see that the thermal quark contribution $F_q$ inhibits and softens chiral symmetry restoration, while the meson contribution $V_\sigma^{\rm ren}$ favors and sharpens it.

\subsection{Fit to lattice data}\label{sec:joy:lattice}

%%%%%%%%%%%%%%%%%%%%%%%%
\begin{figure}
\includegraphics[width=\linewidth]{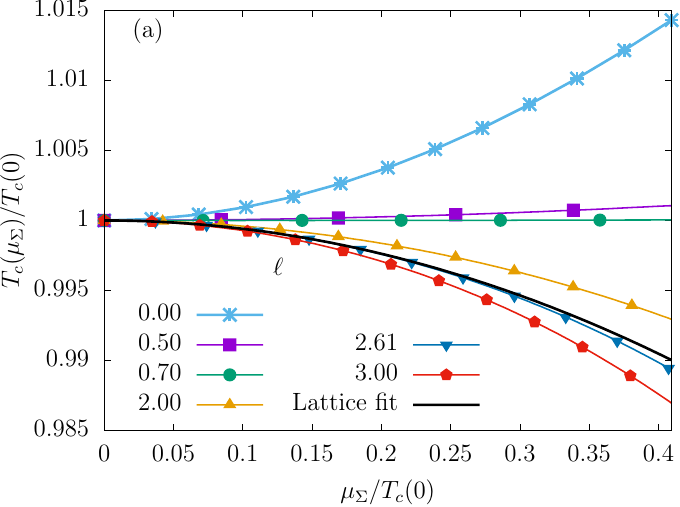} \\
\includegraphics[width=\linewidth]{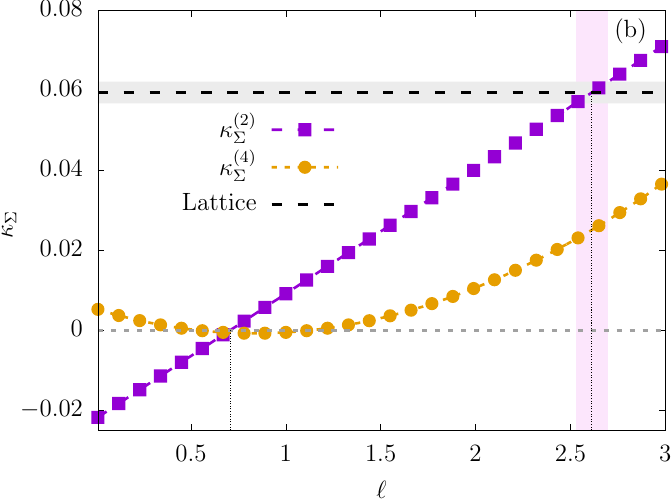}
\caption{(a) Ratio $T_c(\mu_\Sigma) / T_c(0)$ between the pseudocritical temperatures at finite and vanishing spin potentials, as a function of the ratio $\mu_\Sigma / T_c(0)$, for various values of $\ell$. The black line represents the first two terms in the series~\eqref{eq:Tc_real}, up to second order, with $\kappa_\Sigma^{(2)} = \kappa_\Sigma^\psi \simeq 0.06$ fixed from the lattice QCD value reported in Ref.~\cite{Braguta:2025ddq}.
(b) Curvature parameters $\kappa_\Sigma^{(2)}$ and $\kappa_\Sigma^{(4)}$ as functions of the renormalization parameter $\ell$.
}
\label{fig_pd_small}
\end{figure}
%%%%%%%%%%%%%%%%%

We are now in a position to make the connection between the effective model study presented in this paper and the lattice QCD analysis reported in Ref.~\cite{Braguta:2025ddq}. Due to the infamous sign problem, that work considered an imaginary spin potential, $\mu_\Sigma^I = -i \mu_\Sigma$, and found that the pseudocritical transition temperature, $T_{\rm pc}(\mu_\Sigma^I)$, increases quadratically with $\mu_\Sigma^I$. This behavior is characterized by the curvature $\kappa_\Sigma$, as defined in Ref.~\cite{Braguta:2025ddq}:
\begin{equation}
 T_c(\mu_\Sigma^I) \simeq T_c(0) \left[1 + \kappa_\Sigma \left(\frac{\mu_\Sigma^I}{T}\right)^2\right],
 \label{eq:Tc_im}
\end{equation}
valid at small $\mu_\Sigma^I / T$. In QCD, the above formula can refer to the pseudocritical temperature of the deconfinement or chiral transitions, when $\kappa_\Sigma \to \kappa_\Sigma^L$ or $\kappa_\Sigma^\psi$, respectively. In this paper, we focus on the latter coefficient, extrapolated to physical pseudoscalar and vector meson masses, which was reported in Ref.~\cite{Braguta:2025ddq} as
\begin{equation}
 \kappa_\Sigma^\psi \simeq 0.0595(27)\,,
 \label{eq:kappaS_lattice}
\end{equation}
where the number in the parentheses indicates the numerical uncertainty in the last two digits corresponding to a combined statistical and systematic error of Monte Carlo simulations.

Moving now to the case of a real spin potential, Eq.~\eqref{eq:Tc_im} can be analytically continued as follows:
\begin{equation}
 T_c(\mu_\Sigma) \simeq T_c(0) \left[1 - \kappa_\Sigma^{(2)} \left(\frac{\mu_\Sigma}{T_c(0)}\right)^2 - \kappa_\Sigma^{(4)} \left(\frac{\mu_\Sigma}{T_c(0)}\right)^4\right],
 \label{eq:Tc_real}
\end{equation}
where $\kappa_\Sigma^{(2)}$ is expected to retain its value in Eq.~\eqref{eq:kappaS_lattice}, and we included the fourth-order term characterized by $\kappa_\Sigma^{(4)}$ for completeness. Equation~\eqref{eq:Tc_real} indicates a decreasing dependence of $T_c$ on $\mu_\Sigma$. As shown in Fig.~\ref{fig_PT_comp}, this behavior is consistent with the effect of the renormalized meson potential, $V_\sigma^{\rm ren}$ (visible when $\ell = 1$), and opposite to the effect of the thermal quark contribution, $F_q$ (visible when $\ell = 0$).

The above discussion indicates that the parameter $\ell$ allows us to change the regimes between the thermal quark-dominated behavior ($\ell = 0$) to the renormalized meson-dominated behavior (at $\ell \gtrsim 1$). In panel (a) of Fig.~\ref{fig_pd_small}, we represented the ratio $T_c(\mu_\Sigma) / T_c(0)$ with respect to $\mu_\Sigma / T_c(0)$ for various values of $\ell$. The points marked ``Lattice fit'' correspond to the values of $T_c(\mu_\Sigma) / T_c(0)$ computed using Eq.~\eqref{eq:Tc_real} up to the second order, using the curvature $\kappa_\Sigma^{(2)}=\kappa_\Sigma^\psi \simeq 0.06$, compatible with the lattice result~\eqref{eq:kappaS_lattice} within the statistical uncertainty of the numerical simulations. As expected, when $\ell$ is small, the pseudocritical temperature increases with $\mu_\Sigma$, in contrast to the lattice result. The critical value of $\ell$ corresponding to vanishing curvature $\kappa_\Sigma^{(2)}$ is $\ell_0 \simeq 0.70$, while for $\ell > \ell_0$, the curvature $\kappa_\Sigma^{(2)}$ becomes positive, as required by the lattice result in Eq.~\eqref{eq:kappaS_lattice}.

Panel (b) of Fig.~\ref{fig_pd_small} shows the fitted values of $\kappa_\Sigma^{(2)}$ and $\kappa_\Sigma^{(4)}$ as functions of the renormalization parameter $\ell$. The curvature $\kappa_\Sigma^{(2)}$ shows a monotonically-increasing trend. At $\ell = 0$, one finds $\kappa_\Sigma^{(2)} \simeq -0.022$ and $\kappa_\Sigma^{(4)} \simeq 0.005$. The value $\kappa_\Sigma^{(2)} = 0$ is reached at $\ell = 0.70$, as indicated in panel (a). The range given in Eq.~\eqref{eq:kappaS_lattice}, shown with a thin, blue horizontal band around $\kappa_\Sigma^\psi = 0.0595$, corresponds to the following interval of the parameter $\ell$:
\begin{equation}
 \ell = 2.61(8)
\end{equation}
which is shown as a thin, vertical purple band around the value $\ell = 2.61$.
Our fitting procedure also allows us to predict the value for the quartic coefficient:
\begin{align}
    \kappa_\Sigma^{(4)} = 0.0252(24)\,,
    \label{eq_kappa_4}
\end{align}
which was not reported in Ref.~\cite{Braguta:2025ddq}.
For definiteness, we will employ this central value of $\ell = 2.61$ as the ``lattice-compatible'' $\ell$ in what follows, which corresponds to the rounded value $\kappa_\Sigma^\psi = 0.06$. The fit was performed in the range $\mu_\Sigma / T_c(0)\in [0,0.1]$.

\subsection{Pressure and spin density} \label{sec:joy:spin}

\begin{figure}
\includegraphics[width=\linewidth]{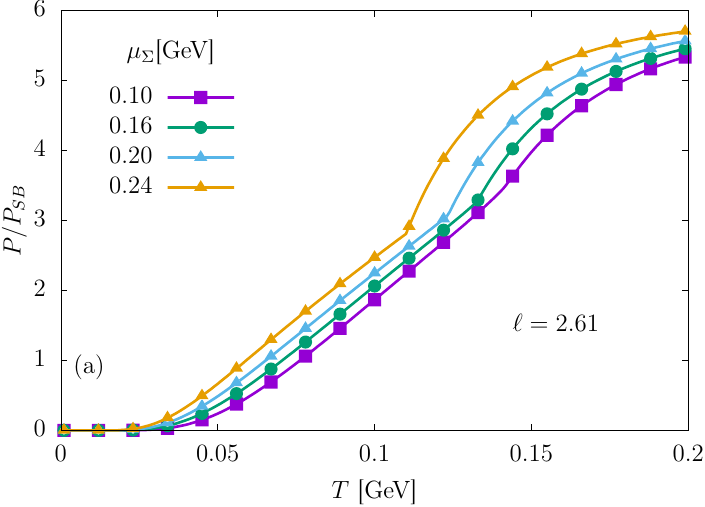}\\
\includegraphics[width=\linewidth]{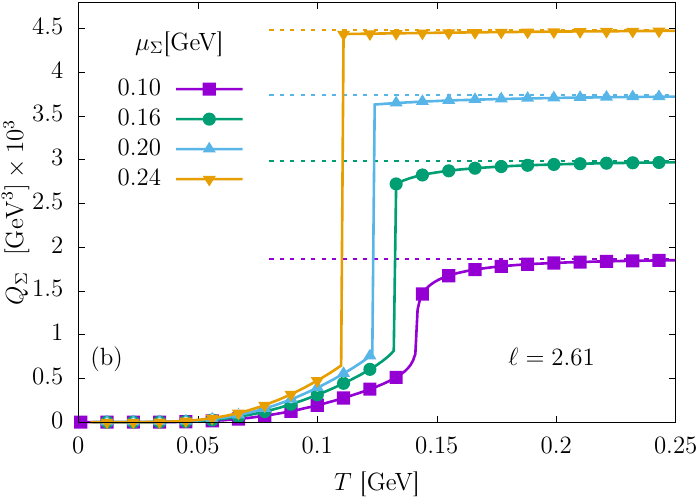}
\caption{(a) Normalized pressure and (b) Spin density as a function of temperature for different values of $\mu_\Sigma$. The dashed lines in panel (b) represent the asymptotic form of $Q_\Sigma^{\rm ZP}$ in the chirally-restored phase, given in Eq.~\eqref{eq:QZP_as}.
}
\label{fig_PQs}
\end{figure}

We now consider the behavior of the system pressure at finite $T$, $P = -F = -V_\sigma^{\rm ren} - F_q$.
In Fig.~\ref{fig_PQs}(a), we show $P$ as a function of temperature for various values of $\mu_\Sigma$. The pressure is normalized by the Stefan-Boltzmann expression, $P_{\rm SB} = 7 N_c N_f \pi^2 T^4/ 180$, corresponding to the leading-order contribution to the large temperature expansion, shown in Eq.~\eqref{eq:PFD_highT}. As shown in Fig.~\ref{fig_PQs}(a), $P$ becomes independent of $\mu_\Sigma$ at large temperatures. At vanishing temperature, $P$ vanishes in the chirally-broken phase, whenever $\mu_\Sigma \sqrt{\ell} \lesssim 0.5$ GeV (see discussion in Sec.~\ref{sec:vac_ren:phase}). Above this threshold value, the zero-point pressure exhibits a quadratic dependence on $\mu_\Sigma$, with the thermal contribution becoming dominant when $T^2 \gtrsim 0.09 g f_\pi \mu_\Sigma \sqrt{\ell}$.

%%%%%%%%%%%%%%%%%
\begin{figure*}
%\begin{figure}[!htb]
\centering
\begin{tabular}{cc}
 \includegraphics[width=0.49\linewidth]{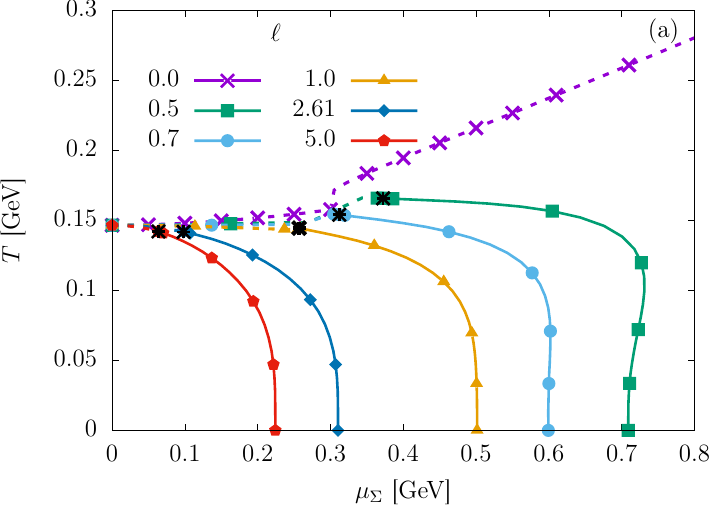} &
 \includegraphics[width=0.49\linewidth]{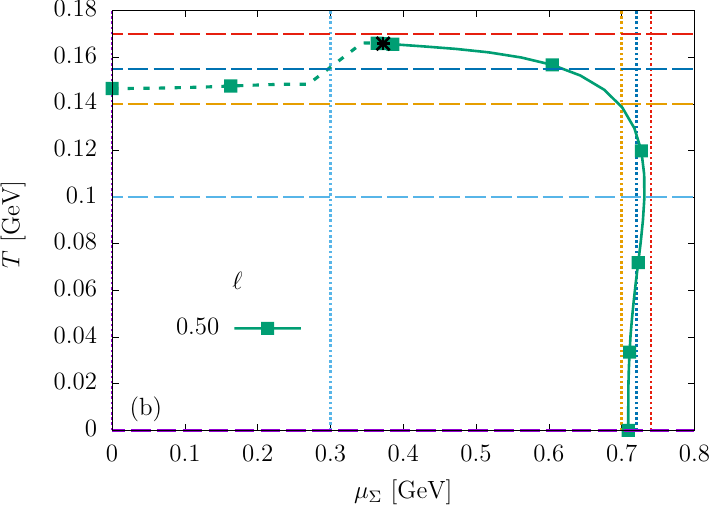} \\
 \includegraphics[width=0.49\linewidth]{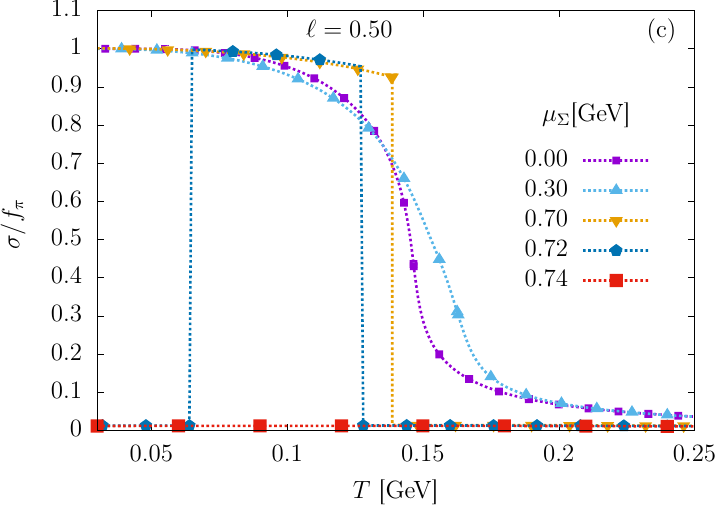} &
 \includegraphics[width=0.47\linewidth]{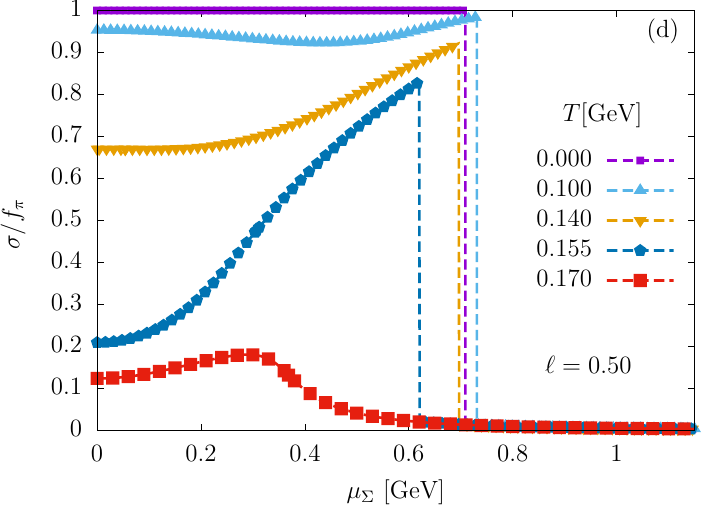}

\end{tabular}
\caption{(a) Phase diagram of the chiral transition, in the $(\mu_\Sigma, T)$ plane, for various values of the parameter $\ell$. The dashed line represents a crossover phase transition, while the solid lines depict first-order phase transitions, separated by the critical points denoted by black stars. (b) Phase diagram for $\ell=0.5$. (c) Order parameter $\sigma/f_\pi$ as a function of temperature for different values of $\mu_\Sigma$, at the values of $\mu_\Sigma$ corresponding to the dotted vertical lines in panel (b); (d) Order parameter $\sigma/f_\pi$ as a function of $\mu_\Sigma$ for the values of $T$ corresponding to the horizontal dashed lines in panel (b).
}
\label{fig_PD_all_l}
\end{figure*}
%%%%%%%%%%%%%%%%%%%%

In the case of the spin charge, $Q_\Sigma = -\partial F / \partial \mu_\Sigma = Q^{\rm ZP}_\Sigma + Q^q_\Sigma$,
the zero-point contribution $Q^{\rm ZP}_\Sigma$ is shown in Eq.~\eqref{eq:QZP}, while the thermal quark contribution can be obtained from $F_q$, shown in Eq.~\eqref{eq:F}:
\begin{multline}
    Q_\Sigma^q = \frac{N_c N_f T}{\pi^2}\sum_{s} s \int_{g\sigma}^\infty dE\, p \left[\frac{\beta \tilde{E}_s\, \mathrm{sgn}(\tilde{E}_s)}{1+e^{\beta |\tilde{E}_s|}}\right.\\
    \left. -\ln \left(1+e^{-\beta |\tilde{E}_s|}\right)\right].
\end{multline}
As discussed in Sec.~\ref{sec:vac_ren:observables}, $Q_\Sigma^{\rm ZP}$ vanishes in the chirally broken phase, when $\sigma = f_\pi$. On the other hand, in the chirally-restored phase,
\begin{align}
	Q^q_\Sigma \simeq \frac{N_c N_f}{8\pi^2} g^2 \sigma^2 \biggl(1 - 2\gamma_E+ 2 \ln \frac{|\mu_\Sigma|}{\pi T}\biggr) \mu_\Sigma + O(\sigma^4)\,,
\end{align}
becomes negligible, since $\sigma = O(T^{-2})$ is small. Figure~\ref{fig_PQs}(b) confirms that after the chiral restoration, $Q_\Sigma$ is dominated by the zero-point contribution, $Q_\Sigma^{\rm ZP}$, shown in Eq.~\eqref{eq:QZP}. Before the chiral restoration, $Q_\Sigma^{\rm ZP} \simeq 0$ and the buildup before the phase transition is largely due to the quark thermal contribution, $Q_\Sigma^q$.

\section{Phase diagram}
\label{sec:res}

In this section, we focus on the phase diagram corresponding to the chiral phase transition in the presence of a spin potential, represented in the $(\mu_\Sigma, T)$ plane.

Panel (a) of Fig.~\ref{fig_PD_all_l} shows the phase diagram corresponding to values of $\ell$ between $0$ and $5$. For $\ell = 0$, the transition remains of crossover type for any $\mu_\Sigma$ and the transition temperature increases with $\mu_\Sigma$. For any finite $\ell$, the discussion in Sec.~\ref{sec:vac_ren:phase} reveals that, at $T = 0$, the system undergoes a first-order phase transition when $\mu_\Sigma \sqrt{\ell} \simeq 0.50$ GeV. Since the transition at $\mu_\Sigma = 0$ is always of crossover type, this implies that for any $\ell > 0$, the phase diagram exhibits a critical point (shown by the dark star symbols).
Furthermore, the requirement that the system undergoes a phase transition at $T = 0$ forces the transition line to bend downwards. Thus, the renormalized meson potential $V_\sigma^{\rm ren}$ competes with the effect of the thermal quark contribution, $F_q$, as can be seen in the case $\ell = 0.5$. When the curvature $\kappa_\Sigma^{(2)}$ changes sign, i.e. for $\ell \gtrsim 0.70$, the renormalized meson potential becomes dominant and the transition temperature stops increasing with $\mu_\Sigma$, for small values of $\mu_\Sigma$. Even when $\kappa_\Sigma^{(2)}$ is positive, our numerical calculations reveal a non-monotonic dependence of the transition temperature on $\mu_\Sigma$. The temperature monotonically decreases with $\mu_\Sigma$ when $\ell>1.22$, as seen in Fig.~\ref{fig_PD_all_l}(a).

For any $0<\ell\lesssim 1.22$, for example  $\ell = 0.5$, shown in Fig.~\ref{fig_PD_all_l}(b), we notice that the competition between $V_\sigma^{\rm ren}$ and $F_q$ leads to a nonmonotonic behavior of the transition line. On the right side of the plot, a vertical crossing of the transition line at $0.706$ GeV $\lesssim \mu_\Sigma \lesssim 0.740$ GeV drives the system from the chirally-restored phase (small $T$) to the chirally-broken phase and then back to the chirally restored phase, via first-order phase transitions. This property is illustrated in panel (c) of Fig.~\ref{fig_PD_all_l}, where we can see the behavior of $\sigma / f_\pi$ with respect to temperature for the values of $\mu_\Sigma$ corresponding to the vertical dotted lines shown in panel (b) of Fig.~\ref{fig_PD_all_l}.

Similarly, in the top region, keeping $T$ fixed for the narrow temperature range $T_c(0) \simeq 0.15$ GeV $\lesssim T \lesssim 0.17$~GeV and varying $\mu_\Sigma$, one drives the system from the chirally-restored phase to the broken phase through a crossover transition, and back to the restored phase through a first-order transition. This behaviour is illustrated in panel (d) of Fig.~\ref{fig_PD_all_l}, where $\sigma / f_\pi$ is represented with respect to $\mu_\Sigma$ for the values of $T$ corresponding to the horizontal dashed lines shown in panel~(b) of Fig.~\ref{fig_PD_all_l}.
\begin{figure}
%\centering
%\begin{tabular}{cc}
\includegraphics[width=\linewidth]{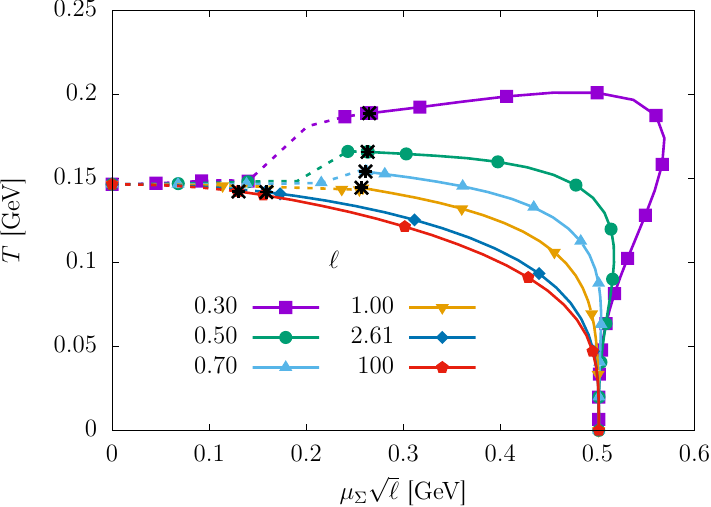}
%(a) & (b) \\[1mm]
%\end{tabular}
\caption{Phase diagram in $(\mu_\Sigma\sqrt{\ell}, T)$ plane for various values of the parameter $\ell$.
For each value of the parameter $\ell$, the crossover transition (the dashed line) is separated from the first-order phase transition (the solid line) by the critical endpoint (the black star).
}
\label{fig_pd_diagonal_scaled}
\end{figure}

%%%%%%%%%%%%%%%%%%%%
\begin{figure}
%\centering
%\begin{tabular}{cc}
\includegraphics[width=\linewidth]{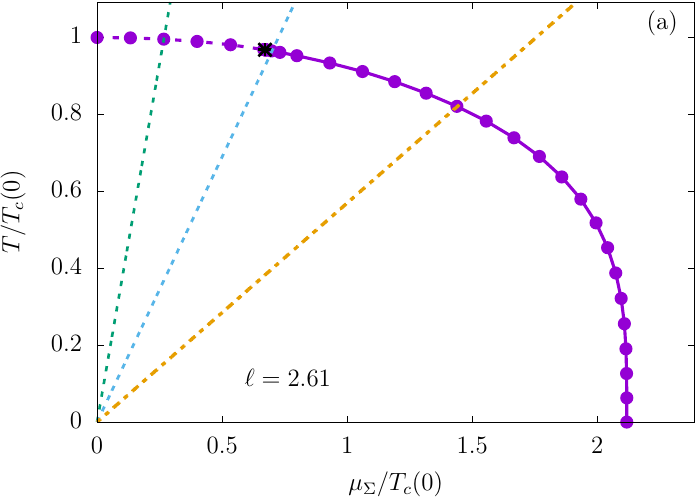} \\
\includegraphics[width=\linewidth]{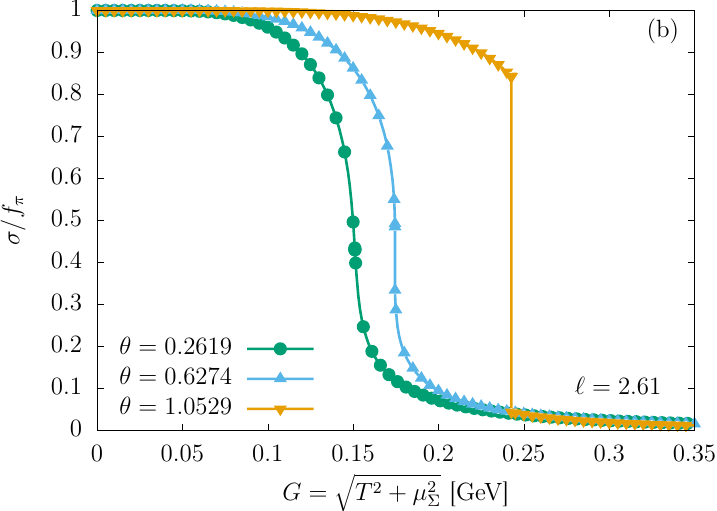}
%(a) & (b) \\[1mm]
%\end{tabular}
\caption{(a) Phase diagram in the $(\mu_\Sigma, T)$ plane for $\ell=2.61$, fixed from the lattice results, with dashed lines indicating a crossover, while the first-order phase transition is depicted using a solid line. The black star represents the critical endpoint separating these regimes.
(b) $\sigma/f_\pi$ with respect to $G=\sqrt{T^2+\mu_\Sigma^2}$ for
the angles $\theta$ corresponding to the dotted lines shown in panel (a).
}
\label{fig_pd_diagonal}
\end{figure}
%%%%%%%%%%%%%%%%%%%%%%%%

We now discuss the scaling of the transition line as $\ell$ is increased to arbitrarily large values. The transition temperature $T_c(0)$ at vanishing spin potential represents a common point on the transition line for all values of $\ell$. Similarly, the transition spin potential at vanishing temperature satisfies the scaling $\mu_\Sigma \sqrt{\ell} \simeq 0.5$ GeV. Thus, representing the phase diagram with respect to the $(T, \mu_\Sigma \sqrt{\ell})$ axes reveals that the transition line converges as $\ell$ is increased to a unique line, as shown in Fig.~\ref{fig_pd_diagonal_scaled}. When $\ell \gg 1$ but $\ell \mu_\Sigma^2$ is kept finite, the spin potential becomes vanishingly small and the thermal quark contribution can be evaluated at vanishing $\mu_\Sigma$. The only remaining spin-dependent contribution is the renormalized meson potential, which contains the quadratic term $\ell \mu_\Sigma^2$.

Finally, panel (a) of Fig.~\ref{fig_pd_diagonal} shows the phase diagram for the value $\ell = 2.61$, corresponding to the curvature parameter $\kappa_\Sigma^{(2)} \simeq 0.06$ computed in lattice QCD (see discussion in Sec.~\ref{sec:joy:lattice}). The vertical ($T$) and horizontal ($\mu_\Sigma$) axes are scaled by the pseudocritical temperature $T_c(0)$ at vanishing spin potential. At low $\mu_\Sigma$, the transition is of crossover type, indicated as a dotted segment. The crossover is separated from the region of first-order phase transition (the solid line) by a critical endpoint (CEP, denoted by the black star) at
\begin{align}
    {(T, \mu_{\Sigma})}_{\rm CEP} \simeq ( 0.142,0.098) \, {\rm GeV}\,.
    \label{eq:T_cep}
\end{align}

As $\mu_{\Sigma}$ increases further, the line of the first-order phase transition hits the $T=0$ axis at the critical point
\begin{align}
	{(T, \mu_{\Sigma})}_{\rm c} \simeq (0, 0.310) \ {\rm GeV}\,.
    \label{eq:Tc}
\end{align}
At higher values of the spin potential, $\mu_{\Sigma} > \mu_{\Sigma,{\rm c}}$, the thermodynamic ground state in this model corresponds to a chirally restored phase for all temperatures.

In panel $(b)$ of Fig.~\ref{fig_pd_diagonal}, we show the behavior of the control parameter $\sigma / f_\pi$ with respect to the search parameter $G = \sqrt{T^2 + \mu_\Sigma^2}$, for three values of the angle $\theta = \tan^{-1}(\mu_\Sigma / T)$, indicated by the dotted lines in panel (a) of Fig.~\ref{fig_pd_diagonal}. The angles $\theta \simeq 0.26$ rad and $\theta \simeq 1.05$ rad lie deep in the crossover and first-order regions, respectively, while $\theta \simeq 0.63$ rad is in the first-order region, just after the critical point.

\section{Conclusions}
\label{sec:conclusion}
%%%%%%%%%%%%%%%%%%%%%%%%%%%%%%%%%%%%%%%%%%%%%%%%%%%%%%%%%%%%%%%%%%%%%%%%%%%%%%%%%%%%%%%%%%%%%%%%%%%%%%%%%%%%%%%%%%%%%%%%%%%%%%%%%%%%%%%%%%%%%%%%

In this work, we investigated the impact of a finite spin density on the QCD phase diagram, with spin polarization characterized by the spin potential, $\mu_\Sigma$. The spin potential represents the energy difference associated with adding, to a system of relativistic fermions, a particle of one spin orientation relative to the other.\footnote{See the discussion on the precise thermodynamical interpretation of the spin potential at the end of Subsec.~\ref{subsec:quark}.} It quantifies the preference for a particular spin state and provides a natural thermodynamic variable that incorporates the spin polarization into the thermodynamic description of the system.

To study the effect of the spin polarization on the thermodynamics of strongly interacting matter, we employed the Linear Sigma model coupled to quarks, ${\rm LSM}_q$, which describes both quark and meson degrees of freedom. We worked in the mean-field approximation, in which fluctuations of the bosonic fields around their ground-state values are neglected, and only the contributions arising from the fermion determinant are taken into account. This approach simplifies the treatment of the thermodynamics of the model, while preserving the essential features of quark dynamics.

We demonstrate that the presence of a finite spin potential, $\mu_\Sigma \neq 0$, leads to a nontrivial dependence of the energy of a fermion quasiparticle on its momentum~\eqref{eq:Energy_dispersion_mus_final}. The particularities of the fermion energy become especially important in the coupling regime where the spin potential dominates over the mass of the fermion quasiparticle, $\tfrac{1}{2}|\mu_\Sigma| > g|\sigma|$. In this domain of parameters --- which is especially important in the transition region between the chirally broken and chirally restored phases --- the mean-field analysis of the thermodynamic ground state requires a more careful treatment to properly capture the modifications induced by a finite spin density. We discussed these subtle questions in our work.

The nontrivial structure of the energy of the fermion quasiparticle at finite spin potential has several important consequences. First of all, one immediately finds that the zero-point (vacuum) free energy generated by the fermion loop exhibits an ultraviolet-divergent contribution that explicitly depends on $\mu_\Sigma$. This feature appears to be in sharp contrast with the standard field-theoretical paradigm, in which the ultraviolet divergencies are only associated with vacuum fluctuations of the fields that are independent of the characteristics of matter, such as temperature and/or chemical potentials. Therefore, in the standard case, the ultraviolet divergencies --- both in the presence or in the absence of matter --- can systematically be incorporated through a well-defined loop-by-loop renormalization of the bare couplings of the theory by matter-independent vacuum effects~\cite{kapusta_gale_2006}. Therefore, one arrives at an intuitive picture that separates vacuum and matter effects: the thermal and quantum fluctuations of matter generate a finite correction to thermodynamic quantities, while the divergent quantum contributions coming from the vacuum can be systematically isolated and treated within a consistent renormalization scheme.

The explicit dependence of the divergent terms on matter-specific properties challenges the standard renormalization procedure, which is employed to absorb ultraviolet divergences into redefined couplings of the original model. This observation highlights the importance of using a consistent framework that disentangles medium effects---arising from $\mu_\Sigma$-dependent divergent contributions to the
total energy---from the genuine divergent terms associated with the zero-point (vacuum) free energy. In this work, we adopt the general strategy of the medium-separation scheme, previously developed for quark matter in the presence of the chiral chemical potential~\cite{Farias:2016let, Azeredo:2024sqc} and the isospin chemical potential~\cite{Avancini:2019ego, Lopes:2021tro, Ayala:2023cnt}. In both these cases, the corresponding chemical potentials produce nontrivial modifications of the quasiparticle dispersion relation, analogous to the effects generated by the spin potential in
Eq.~\eqref{eq:Energy_dispersion_mus_final}.

The renormalization procedure introduces a free model parameter, $\ell$, which encodes an interplay of vacuum fluctuations with those of spin-polarized matter. Although this parameter cannot be determined from field-theoretical considerations, its value can be constrained by matching the curvature of the
chiral phase transition predicted within the LSM$_q$ to nonperturbative results of lattice QCD simulations~\cite{Braguta:2025ddq}. Due to the well-known sign problem, lattice computations are restricted to the regime of small spin polarizations, where such a comparison is feasible. Once fixed, our approach allows us to make a consistent extrapolation of lattice results beyond this limited domain and, therefore,
extend the description of the chiral phase transition to the full $(\mu_\Sigma, T)$ plane.

As a result, we obtain the phase diagram shown in Fig.~\ref{fig_pd_diagonal}(a). The phase diagram contains a chirally broken phase separated from the chirally restored phase by a transition line. This chiral transition is a crossover in a small domain of values of the spin potential $\mu_\Sigma$. For $|\mu_\Sigma| \ll T_c(0)$, the transition line can be expressed via the series~\eqref{eq:Tc_real}, where the curvature of the quadratic term~\eqref{eq:kappaS_lattice} has already been found in the lattice simulations~\cite{Braguta:2025ddq}, while the value of the quartic term, that has not been determined numerically, is predicted by our analytical model in Eq.~\eqref{eq_kappa_4}.

The transition becomes of first order at the critical endpoint given in Eq.~\eqref{eq:T_cep}. The transition-line continues at higher spin potentials and intersects the $T = 0$ axis at another endpoint~\eqref{eq:Tc}. At higher spin potentials, the spin-polarized matter resides in the chirally restored phase, regardless of temperature.

Our investigations revealed several non-intuitive properties of the QCD matter at finite spin potential $\mu_\Sigma$. First, the spin potential does not introduce a Fermi level while leading to a non-trivial modification to the energy dispersion relation~\eqref{eq:Energy_dispersion_mus_general}. The first-order phase transition encountered at vanishing temperature, $T = 0$, is driven by the renormalized meson potential. Second, the thermal quark contribution to the thermodynamic potential becomes independent of $\mu_\Sigma$ in the high-temperature, chirally-restored phase. Here, the polarization effects on the QCD matter are entirely governed by the renormalized meson potential. Third and finally, the spin polarization saturates with respect to increasing the temperature, being given by a temperature-independent term derived from the renormalized meson potential. In the chirally-restored phase, the thermal quark contribution to the spin polarization becomes negligible.

The emergence of the spin degree of freedom as a thermodynamical variable opens a new avenue of research of QCD thermodynamics, including the influence of spin polarization on the phase diagram of QCD, which could potentially include the emergence of new, spin-promoted phases of QCD. Within the current setup, one possible extension would be to explore the deconfinement phase transition properties, using the Polyakov loop extended LSM$_q$ (PLSM$_q$) model.

{\it Note added.} While this paper has been in the final stages of preparation, a similar study has emerged in a generalization of a Polyakov–loop enhanced Nambu–Jona-Lasinio model in Ref.~\cite{Farias:2025vss} (see also Ref.~\cite{Kiefer:2025xdp}). Although the predicted phase diagram qualitatively agrees with our result in Fig.~\ref{fig_pd_diagonal}(a), there are also two important differences: contrary to Ref.~\cite{Farias:2025vss}, our approach not only allows us to precisely match the curvature of the chiral transition to first-principle lattice data at small polarization, but also, as a corollary, leads to the critical end point (CEP) occurring at a substantially lower value of the spin potential~\eqref{eq:T_cep}.

\begin{acknowledgments}
The authors are grateful to Dr. Nyx Shiva for useful comments on the manuscript. This work was funded by the EU’s NextGenerationEU instrument through the National Recovery and Resilience Plan of Romania - Pillar III-C9-I8, managed by the Ministry of Research, Innovation and Digitization, within the project entitled ``Facets of Rotating Quark-Gluon Plasma'' (FORQ), contract no.~760079/23.05.2023 code CF 103/15.11.2022.
\end{acknowledgments}

\section*{Data availability}

The data that support the findings of this article are openly available~\cite{singha_2025_Zenodo}.

\appendix

\section{Dimensional Regularization}\label{app:DR}

In the context of zero-point renormalization within the standard LSM$_q$ model, we introduced the integral $I_1$ in Eq.~\eqref{eq:I1_def}. Additionally, while addressing the axial-vector interaction term, we introduced two more integrals, $I_2$ and $I_3$ in Eqs.~\eqref{eq:I2_def},\eqref{eq:I3_def}. Below, we present the evaluations of these integrals using dimensional regularization. We also discuss the computation of the finite remainder $\delta I_\Sigma$, introduced in Eq.~\eqref{eq:IS_def}.

Under the dimensional regularization prescription, we represent $d$-dimensional momentum integrals as
\begin{align}
    \int\!\!\frac{d^dp}{(2\pi)^d} \to \int\!\!\frac{d\Omega_d}{(2\pi)^d}\int\!\!dp~p^{d-1}.
\end{align}
The angular integral can be computed as
\begin{equation}
 \int\!d\Omega_d = \frac{2\pi^{d/2}}{\Gamma(d/2)}.
    \label{eq:pzn_to_pn}
\end{equation}
For future convenience, we note that the integral of even powers of $p^z$ can be obtained as
\begin{equation}
\int\! d\Omega_d \frac{p_z^2}{p^2} = \frac{1}{d} \int\!d\Omega_d, \quad
\int\! d\Omega_d \frac{p_z^4}{p^4} = \frac{3}{d(d+2)} \int\!d\Omega_d.
\end{equation}
To cope with divergent terms, the dimension of the momentum integral is replaced by $d\to 3-2\epsilon$, with $\epsilon$ a regularization parameter to be taken to $0$ at the end of the calculation. To preserve the dimensionality of the expression, a renormalization energy scale $M$ is employed as a pre-factor $M^{2\epsilon}$. Within the minimal subtraction scheme, one can then express
\begin{align}
    \int\!\frac{d^dp}{(2\pi)^d} f(\mathbf{p}) \to
    \frac{M^{2\epsilon}}{(2\pi)^{3 - 2\epsilon}} \int d\Omega_{3 - 2\epsilon} \int_0^\infty dp\, p^{2-2\epsilon} f(\mathbf{p}).
    \label{eq:dreg_ddp}
\end{align}
The angular integral is evaluated by analytical continuation:
\begin{equation}
 \int d\Omega_{3 - 2\epsilon} = \frac{2\pi^{\frac{3}{2}-\epsilon}}{\Gamma\left(\frac{3}{2}-\epsilon\right)}.
\end{equation}
The next steps involve evaluating the integrals and expanding the results in powers of $\epsilon$. The $1/\epsilon$ contributions represent the divergent terms that must be removed by suitable counterterms.

Applying the dimensional regularization to the integrals $I_1$, $I_2$ and $I_3$, given in Eqs.~\eqref{eq:I1_def}, \eqref{eq:I2_def} and \eqref{eq:I3_def} of the main text, we obtain
\begin{subequations}
\begin{align}
 I_1&= \frac{2\pi^{\frac{3}{2}-\epsilon} M^{2\epsilon}}{(2\pi)^{3-2\epsilon}\Gamma\left(\frac{3}{2}-\epsilon\right)}\int\limits_0^\infty dp\, p^{2-2\epsilon} \left(p^2+g^2\sigma^2\right)^{\frac{1}{2}},  \\
 I_2&= \frac{\pi^{\frac{3}{2}-\epsilon} M^{2\epsilon} (1 - \epsilon)}{4(2\pi)^{3-2\epsilon}\Gamma\left(\frac{5}{2}-\epsilon\right)}\int\limits_0^\infty \frac{dp\, p^{4 - 2\epsilon}}{(p^2+g^2\sigma^2)^{3/2}}, \\
 I_3 &= \frac{\pi^{\frac{3}{2}-\epsilon} M^{2\epsilon} (1 - \epsilon)}{64(2\pi)^{3-2\epsilon}\Gamma\left(\frac{7}{2}-\epsilon\right)}
 %\int\limits_0^\infty \frac{dp\, p^{4 - 2\epsilon}[\epsilon p^2 + 2(5 - 2\epsilon) g^2 \sigma^2]}{(p^2+g^2\sigma^2)^{7/2}}.
 \left[\epsilon \int\limits_0^\infty \frac{dp\, p^{4-2\epsilon}}{(p^2 + g^2 \sigma^2)^{5/2}} \right. \nonumber\\
 & \left. + 5 g^2 \sigma^2(2 - \epsilon) \int\limits_0^\infty \frac{dp\, p^{4-2\epsilon}}{(p^2 + g^2 \sigma^2)^{7/2}}\right].
\end{align}
\end{subequations}
The integrals with respect to $p$ can be performed using the formula:
\begin{multline}
 \int\limits_0^\infty \frac{dp\, p^\alpha}{(p^2 + g^2 \sigma^2)^{\beta/2}} =
 \frac{(g\sigma)^{\alpha - \beta + 1}}{2\Gamma(\beta / 2)} \\\times
 \Gamma \left(\frac{\alpha + 1}{2}\right)
 \Gamma\left(\frac{\beta - \alpha - 1}{2}\right),
\end{multline}
leading to
\begin{subequations}
\begin{align}
 I_1 &= -\frac{\pi^{1-\epsilon} M^{2\epsilon} (g\sigma)^{4-2\epsilon}}{2(2\pi)^{3-2\epsilon}} \Gamma(\epsilon - 2), \\
 I_2 &= -\frac{\pi^{1-\epsilon} M^{2\epsilon} (g\sigma)^{2-2\epsilon}}{4(2\pi)^{3-2\epsilon}}\Gamma(\epsilon), \\
 I_3 &= \frac{\pi^{1-\epsilon} M^{2\epsilon} (g\sigma)^{-2\epsilon}(1 - \epsilon)}{48(2\pi)^{3-2\epsilon}}\Gamma(\epsilon+1).
\end{align}
\end{subequations}
Taking the limit $\epsilon \to 0$, we employ, for all non-negative integers $n \geqslant 0$, the following relation:
\begin{equation}
 \Gamma(\epsilon - n) =
 \frac{(-1)^n}{n!} \left[
 \frac{1}{\epsilon} + \psi(n+1)\right] + O(\epsilon),
\end{equation}
with $\psi(z) = d \ln\Gamma(z) / dz$ being the polygamma function. Retaining terms up to $O(\epsilon^0)$, we uncover the results in Eqs.~\eqref{eq:I1_reg} and \eqref{eq:I2I3_reg} presented in the main text.

The remainder $\delta I_\Sigma$, introduced in Eq.~\eqref{eq:IS_def}, is regular both in the IR and in the UV, and reads:
\begin{multline}
 \delta I_\Sigma = \int \frac{d^3p}{(2\pi)^3} \left[\frac{1}{2} \sum_s E^{(s)}_{\bf p} - E_p - \frac{\mu_\Sigma^2}{8} \left(\frac{1}{E_p} - \frac{E_{p_z}^2}{E_p^3}\right) \right.\\
 \left.+ \frac{\mu_\Sigma^4}{128} \left(\frac{1}{E_p^3} - \frac{6 E_{p_z}^2}{E_p^5} + \frac{5 E_{p_z}^4}{E_p^7}\right)\right].
\end{multline}
Due to the axial symmetry of the integrand, it is convenient to perform the integral with respect to cylindrical coordinates, $d^3p \to 2\pi dp^z dp_\perp p_\perp$.
We perform the improper integration with respect to $p_\perp$ as the limit $\int_0^\infty dp_\perp\, f(p_\perp) = \lim_{P_\perp \to \infty} \int_0^{P_\perp} dp_\perp f(p_\perp)$. Using the relations
\begin{align}
 \int_0^{P_\perp} dp_\perp\, p_\perp E_{\bf p}^{(s)} &= \frac{1}{3} \bigl[(E_{\bf P}^{(s)})^3 - |E_{p_z} - s \mu_\Sigma|^3\bigr], \nonumber\\
 \int_0^{P_\perp} dp_\perp \,p_\perp E_p^{2n+1} &= \frac{1}{2n+3} \bigl(E_P^{2n+3} - E_{p_z}^{2n+3}\bigr),
\end{align}
with $E_{\bf P}^{(s)} = \sqrt{P_\perp^2 + (E_{p_z}- s\mu_\Sigma)^2}$ and
$E_P = \sqrt{P_\perp^2 + E_{p_z}^2}$,
as well as the expansions
\begin{gather}
 [E_{\bf P}^{(s)}]^3 \simeq P_\perp^3 + \frac{3P_\perp}{2} (E_{p_z} - s\mu_\Sigma)^2, \nonumber\\
 E_P^3 \simeq P_\perp^3 + \frac{3P_\perp}{2} E_{p_z}^2, \quad
 E_P \simeq P_\perp,
\end{gather}
and $E_P^{-2n-1} \simeq 0$ for all $n \ge 0$, we arrive at
\begin{multline}
 \delta I_\Sigma = -\int_0^\infty \frac{dp^z}{6\pi^2} \\\times
 \left(\frac{1}{2} \sum_{s = \pm \frac{1}{2}} |E_{p_z} - s\mu_\Sigma|^3- E_{p_z}^3 - \frac{3}{4} E_{p_z} \mu_\Sigma^2\right).
\end{multline}
When $E_{p_z} > \frac{1}{2}|\mu_\Sigma|$, the sum over $s = \pm 1/2$ yields $\frac{1}{2} \sum_s |E_{p_z} - s\mu_\Sigma|^3 = E_{p_z}^3 + \frac{3}{4} E_{p_z} \mu_\Sigma^2$, leading to an exact cancellation of the integrand. Therefore, $\delta I_\Sigma$ receives non-vanishing contribution only when $E_{p_z} < |\mu_\Sigma|$:
\begin{equation}
 \delta I_\Sigma = -\frac{\theta(\frac{1}{2}|\mu_\Sigma| - g|\sigma|)}{6\pi^2} \int_0^{p_\Sigma} dp^z (\tfrac{1}{2}|\mu_\Sigma| - E_{p_z})^3,
\end{equation}
with $p_\Sigma = \sqrt{\frac{1}{4}\mu_\Sigma^2 - g^2 \sigma^2}$ and we took into account that $E_{p_z} < \frac{1}{2}|\mu_\Sigma|$ implies that $\frac{1}{2}|\mu_\Sigma| > g|\sigma|$. Performing the above integration leads to the result shown in Eq.~\eqref{eq:dIS_result}.

%%%%%%%%%%%%%%%%%%%%%%%%%%%%%%%%%%%%%%%%%%%%%%%%%%%%%%%%%%%%%%%%%%%%%%%%%%%%%%%%%%%%%%%%%%%%%%%%%%%%%%%%%%%%%%%%%%%%%%%%%%%%%%%%%%%%%%%%%%%%%%%%%%%%%%%%%%%%%%%%%%%%%%%%%%%%%%%%%%%%%%%%%%%%%%%%%%%%%%%%%%%%%%%%%%%%%%%%%%%%%%%%%%%%%%%%%%%%

\section{Evaluation of the thermal contribution}
\label{app:sad}

In this appendix, we discuss the transition from the representation of the thermal part of the thermodynamic potential, $F_q$~\eqref{eq:Fqth_sph}, and its derivative $\partial F_q / \partial \sigma$ \eqref{eq:dFds_sph}, expressed with respect to the spin-deformed fermion energy $E_{\mathbf{p}}^{(s)}$ defined in Eq.~\eqref{eq:Energy_dispersion_mus_final}, to the more convenient forms in Eqs.~\eqref{eq:F} and \eqref{eq:dFds}, respectively. We begin in Sec.~\ref{app:sad:energy} by pointing out that the energy $E_{\mathbf{p}}^{(s)}$ can be degenerate with respect to $\mathbf{p}$, in the case when $s\mu_\Sigma > g\sigma$.  Sections~\ref{app:sad:case1} and \ref{app:sad:case2} address separately the cases $g\sigma > s\mu_\Sigma$ (case 1) and $g\sigma < s\mu_\Sigma$ (case 2).\footnote{In the case when $\sigma$ takes a negative value, all instances of $\sigma$ should be replaced by $|\sigma|$.}

\begin{figure}
    \centering\includegraphics[width=0.9\linewidth]{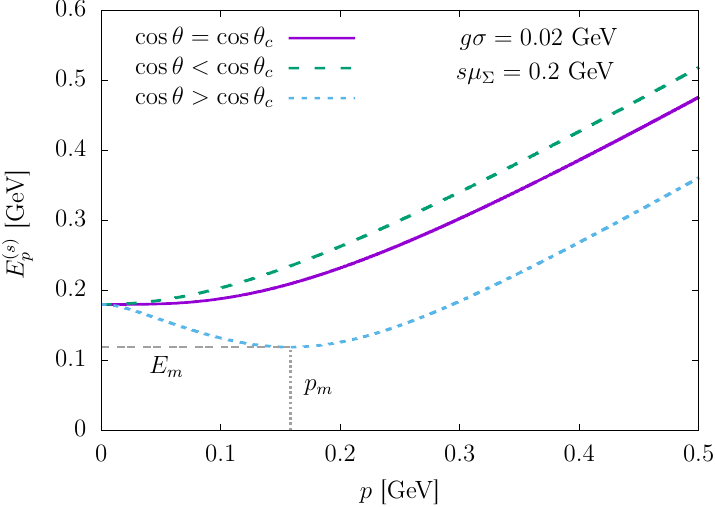}
    \caption{Fermion energy $E_{\mathbf{p}}^{(s)}$ with respect to the momentum $p$, in the case when $ g \sigma< s \mu_\Sigma $, for various values of $\cos\theta = p^z / p$. The critical angle $\theta_c$ is defined in Eq.~\eqref{eq:thetac}. }
    \label{fig:E_theta_phasespace}
\end{figure}

\subsection{Energy branches}\label{app:sad:energy}

Noting the relation \eqref{eq:Energy_dispersion_mus_final} between the fermion energy $E_{\mathbf{p}}^{(s)}$, expressed with respect to the spherical coordinates $(p, \theta, \varphi)$, and the momentum magnitude $p$, we derive
\begin{equation}
 \frac{dE_{\mathbf{p}}^{(s)}}{dp} = \frac{p}{E_{\mathbf{p}}^{(s)}} \left(1 - \frac{s \mu_\Sigma}{E_{p_z}} \cos^2 \theta\right).
 \label{eq:dEdp_mus}
\end{equation}
The above relation allows the integration variable to be changed from $p$ to $E_{\mathbf{p}}^{(s)}$ in
Eqs.~\eqref{eq:Fqth_sph} and \eqref{eq:dFds_sph}:
\begin{subequations}
\label{eq:Fq__dFqds_general}
\begin{align}
       F_q &= -\frac{N_cN_f T}{\pi^2} \sum_{s} \int\limits_{\Gamma_E} dE E \ln\left(1+e^{-\beta E}\right) I^s_{\theta,E},\label{eq:Fq_general}\\
       \frac{\partial F_q}{\partial \sigma} &= \frac{N_cN_f g^2 \sigma}{\pi^2} \sum_{s} \int\limits_{\Gamma_E} \frac{dE}{e^{\beta E} + 1} \left[I^s_{\theta,E}-\delta I^s_{\theta,E}\right],
 \label{eq:dFqds_general}
\end{align}
\end{subequations}
where $\Gamma_E$ represents a suitable contour for the energy integral (see below). In the above, we introduced the angular integrals
\begin{align}
\begin{pmatrix}
 I^s_{\theta,E} \\
 \delta I^s_{\theta,E}
\end{pmatrix}
 = \frac{1}{2} \int\limits_{\Gamma_\theta} \frac{d\cos\theta ~p_s}{E_{p_z}^s - s\mu_\Sigma \cos^2\theta}
 \begin{pmatrix}
  E_{p_z}^s \\ s \mu_\Sigma
 \end{pmatrix},\label{eq:I_s_theta_general}
\end{align}
where
the integration domain $\Gamma_\theta$ depends on the value of the energy $E$, as discussed below. Note that now we view $p \to p_s(E)$ and $E_{p_z} \to E_{p_z}^s = \sqrt{p_s^2 \cos^2\theta + g^2 \sigma^2}$ as functions of the energy $E$ and the spin branch $s = \pm \tfrac{1}{2}$.

Let us now critically analyze the energy structure to get a handle on the energy contour $\Gamma_E$. Setting $p=0$ in Eq.~\eqref{eq:Energy_dispersion_mus_final}, the energy evaluates to $E_0 = |s \mu_\Sigma - g\sigma|$.
On the other hand, setting $dE_{\mathbf{p}}^{(s)}/dp = 0$  at fixed $\theta$ in Eq.~\eqref{eq:dEdp_mus}
implies $E_{p_z} = s\mu_\Sigma \cos^2\theta$. When $s\mu_\Sigma > 0$, this relation
reveals a critical momentum $p_m$,
\begin{align}
 %1 - \frac{s|\mu_\Sigma|}{E_{p_z}} \cos^2\theta = 0 \,\Rightarrow \,
 p_m &= \sqrt{\frac{1}{4}\mu_\Sigma^2 \cos^2\theta - \frac{g^2\sigma^2}{\cos^2\theta}},\label{eq:pm}
 \end{align}
where the energy attains a minimum value $E_m$:
\begin{equation}
  E_m = \sin\theta \sqrt{\frac{1}{4}\mu_\Sigma^2 - \frac{g^2\sigma^2}{\cos^2\theta}}.\label{eq:em}
\end{equation}
The particular structures of $E_0$ and $E_m$ invoke two distinct cases, namely, case 1: $g\sigma > s \mu_\Sigma$ and case 2: $g\sigma < s\mu_\Sigma$. In case 1, $E_m^2 < 0$ and the energy increases monotonically, from $E_0 = g\sigma - s \mu_\Sigma > 0$ to infinity. For case 2, $E_m^2 > 0$ implies that the energy first decreases from $E_0 = s\mu_\Sigma - g\sigma$ to $E_m$, then it increases back to $E_0$ and subsequently to infinity. Furthermore, demanding $p_m^2 > 0$, the expression under the square root of Eq.~\eqref{eq:pm} must be positive, and this is achieved only when $|\cos \theta|$ exceeds a critical value,
\begin{equation}
 \cos\theta_c = \sqrt{\frac{2g |\sigma|}{|\mu_\Sigma|}},
 \label{eq:thetac}
\end{equation}
i.e., when $\cos\theta < -\cos\theta_c$ or $\cos\theta > \cos\theta_c$.

We now have a complete picture of the behaviour of the phase space in the $(E,\theta)$ coordinates,
for case 2: for $|\cos\theta| < \cos\theta_c$, the energy is single valued, increasing from $E_0 = E(p = 0) = \tfrac{1}{2}|\mu_\Sigma| - g\sigma$ to infinity. When $\cos\theta_c < |\cos\theta| \le 1$, the energy first decreases from $E_0$, when $p = 0$, to $E_m$ when $p = p_m$. Note that $E_m = E_0$ when $|\cos \theta| = \cos\theta_c$ ($p_m = 0$) and $E_m = 0$ when $\cos\theta = \pm 1$ ($p_m = \sqrt{\frac{1}{4}\mu_\Sigma^2 - g^2 \sigma^2}$).
From $E_m$, both the energy and the momentum magnitude increase monotonically, first to $E_0$ and $p_M = \sqrt{|\mu_\Sigma|(|\mu_\Sigma| \cos^2\theta - 2g|\sigma|)}$, and then up to infinity. This behavior is illustrated in Fig.~\ref{fig:E_theta_phasespace}. In the remaining two subsections, we will evaluate both $F_q$ and $\partial F_q/\partial\sigma$ within the aforementioned two distinct scenarios.
\subsection{Case 1: \texorpdfstring{$g\sigma > s \mu_\Sigma$}{Effective mass larger than spin potential}}
\label{app:sad:case1}
For case 1, the energy contour follows a monotonic increase from $E_0^1=g\sigma - s \mu_\Sigma$ to infinity. Hence, $\Gamma_E$ in Eqs.~\eqref{eq:Fq__dFqds_general} is replaced by the interval $E_0^1 < E < \infty$. Similarly, the contour $\Gamma_\theta$ in the angular integrals in Eq.~\eqref{eq:I_s_theta_general} is simply $-1 \le \cos\theta \le 1$.

We now compute the angular integrals $I^s_{\theta,E}$ and $\delta I^s_{\theta,E}$. Using the relations $E_{p_z}^s = \sqrt{p_s^2 \cos^2\theta + g^2 \sigma^2}$ and $E^2 = p_s^2 + g^2\sigma^2 + \frac{1}{4}\mu_\Sigma^2 - 2 s \mu_\Sigma E_{p_z}^s$, we can derive
\begin{align}
 \frac{d E_{p_z}^s}{d\cos\theta} &= \frac{p_s \cos\theta}{E_{p_z}^s}\left(\cos\theta \frac{dp_s}{d\cos\theta} + p_s\right), \nonumber\\
 \frac{dp_s}{d\cos\theta} &= \frac{s \mu_\Sigma p_s \cos\theta}{E_{p_z}^s - s \mu_\Sigma \cos^2\theta},\label{eq:dpsdcostheta}
\end{align}
and re-label $p_s\to p$ to further simplify $I^s_{\theta,E}$ and $\delta I^s_{\theta,E}$ as:
\begin{align}
 (I_{\theta,E}^s, \delta I^s_{\theta,E}) = \int\limits_{p^s_0}^{p^s_1} \frac{dp}{\cos\theta} \left(\frac{E_{p_z}^s}{s\mu_\Sigma}, 1\right),
\end{align}
with $E_{p_z}^s = (p^2 + g^2 \sigma^2 + \frac{1}{4}\mu_\Sigma^2 - E^2) / s \mu_\Sigma$ and
\begin{subequations}
\label{eq:ps01}
\begin{align}
 p^s_0 &\equiv p_s(\cos\theta = 0) = \sqrt{E^2 - (g\sigma - s\mu_\Sigma)^2}, \label{eq:ps0}\\
 p^s_1 &\equiv p_s(\cos\theta = 1) = \sqrt{(E + s \mu_\Sigma)^2 - g^2 \sigma^2}.
 \label{eq:ps1}
\end{align}
\end{subequations}
Since $(p_1^s)^2 - (p_0^s)^2 = 2s\mu_\Sigma(E + s\mu_\Sigma - g\sigma)$, $p_1^s > p_0^s$ when $s \mu_\Sigma > 0$ and $p_1^s < p_0^s$ if $s \mu_\Sigma < 0$.

Expressing $\cos\theta$ as
\begin{align}
 \cos\theta &= \frac{p_{s;+} p_{s;-}}{|\mu_\Sigma| p}, \nonumber\\
 p^2_{s;\pm} &= {\rm sgn}(s \mu_\Sigma) \left[p^2 + (g\sigma \pm s \mu_\Sigma)^2 - E^2\right],
\end{align}
we arrive at
\begin{equation}
 (I_{\theta,E}^s, \delta I_{\theta,E}^s) = \int\limits_{p^s_0}^{p^s_1} \frac{dp\, p}{p_{s;+} p_{s;-}}
 \left(\frac{p_{s;+}^2 + p_{s;-}^2}{2 s\mu_\Sigma}, |\mu_\Sigma|\right).
\end{equation}
Since $p_{s;-}^2 = {\rm sgn}(s \mu_\Sigma)[p^2 - (p^s_0)^2]$ and $p_{s;+}^2 = p_{s;-}^2 + 2 g \sigma |\mu_\Sigma|$, we can switch the integration variable to $y = p_{s;-}^2$, leading to
\begin{equation}
 (I_{\theta,E}^s, \delta I_{\theta,E}^s) = \int\limits_0^{Y_s} \frac{dy}{\sqrt{y(y + 2 |\mu_\Sigma| g \sigma)}}
 \left(\frac{y}{|\mu_\Sigma|} + g\sigma, s \mu_\Sigma\right),
\end{equation}
 where $Y_s = |\mu_\Sigma| (E - g\sigma + s \mu_\Sigma)$.
Using the relations
\begin{subequations}
    \begin{align}
 \int\limits_0^{Y_s} \frac{dy (y + a_s)}{\sqrt{y(y+2a_s)}} &= \sqrt{Y_s(Y_s+2a_s)}, \\
 \int\limits_0^{Y_s} \frac{dy}{\sqrt{y(y+2a_s)}} &= \ln \left[1 + \frac{Y_s}{a_s} + \sqrt{\frac{Y_s}{a_s} \left(2 + \frac{Y_s}{a_s}\right)}\right],
\end{align}
\end{subequations}
with $a_s = |\mu_\Sigma| g \sigma$, we arrive at
\begin{align}
 I_{\theta,E}^s &= p^s_1, &
 \delta I_{\theta,E}^s &= s \mu_\Sigma \ln \left(\frac{E + s \mu_\Sigma + p^s_1}{g\sigma}\right).
\end{align}
Replacing $I^s_{\theta,E}$ and $\delta I^s_{\theta,E}$ in Eqs.~\eqref{eq:Fq__dFqds_general}, we eventually obtain
\begin{subequations}
\label{eq:Fq&dFds_case1_aux}
\begin{align}
 F_q &= -\frac{N_cN_fT}{\pi^2} \sum_{s} \int\limits_{g\sigma - s \mu_\Sigma}^\infty dE\, E p_1^s \ln (1 + e^{-\beta E}), \label{eq:Fq_case1_aux}\\
 \frac{\partial F_q}{\partial \sigma} &= \frac{N_cN_f}{\pi^2} g^2 \sigma \sum_{s} \int\limits_{g\sigma - s \mu_\Sigma}^\infty \frac{dE}{e^{\beta E} + 1} \nonumber\\
 & \times \left[p^s_1 - s\mu_\Sigma \ln \left(\frac{E + s\mu_\Sigma + p^s_1}{g\sigma}\right)\right],
 \label{eq:dFds_case1_aux}
\end{align}
\end{subequations}
with $p_1^s = \sqrt{(E + s\mu_\Sigma)^2 - g^2 \sigma^2}$ defined in Eq.~\eqref{eq:ps1}.
Shifting the energy to $E + s\mu_\Sigma \to E$ in Eqs.~\eqref{eq:Fq&dFds_case1_aux} recovers Eqs.~\eqref{eq:suitcaseless} in the main text.

\subsection{Case 2: \texorpdfstring{$g \sigma < s\mu_\Sigma$}{gs}}
\label{app:sad:case2}
As discussed in Sec.~\ref{app:sad:energy}, the fermion energy $E_{\mathbf{p}}^{(s)}$ becomes degenerate when $g \sigma < s\mu_\Sigma$. Separating $F_q = F_q^+ + F_q^-$, with $F_q^{\pm}$ corresponding to $s \mu_\Sigma = \pm \tfrac{1}{2}|\mu_\Sigma|$, we note that the $F_q^-$ contribution is compatible with case 1 discussed in Sec.~\ref{app:sad:case1}. We therefore focus for the remainder of this section on $F_q^+$, which we split as follows:
\begin{equation}
 F_q^+ = F_q^< + F_q^> + F_q^c,
\end{equation}
where the symbols $<$, $>$ and $c$ refer to various integration ranges, as follows:
\begin{equation}
(\Gamma^r_\theta,\Gamma^r_E) =
\begin{cases}
([0,\cos\theta_c],[E_0,\infty)), & r = <, \\
([\cos\theta_c,1],\int\limits_c), & r = c, \\
([\cos\theta_c,1],[E_0,\infty)), & r = >,
\end{cases}
\label{eq:case2_integration_regions}
\end{equation}
where we took into account that the integrands of $I^s_{\theta,E}$ and $\delta I^s_{\theta, E}$ are even with respect to $\cos\theta$ and therefore the integration range runs only over non-negative values of $\cos\theta$.
On the $<$ branch, $\cos\theta < \cos\theta_c$, $p_m^2 < 0$ and the energy is monotonic and increases from $E_0 = \tfrac{1}{2}|\mu_\Sigma| - g\sigma$ to infinity. The $>$ branch corresponds to the region with $p > p_M$ (see end of Sec.~\ref{app:sad:energy}), on which the energy again increases monotonically from $E_0 = \tfrac{1}{2}|\mu_\Sigma| - g\sigma$ to $\infty$. The $r= c$ branch spans the critical region, when $\cos\theta > \cos\theta_c$ and the energy first decreases from $E_0$ to $E_m$ and then increases back to $E_0$.

To get a handle of what happens within the integration regions shown in Eq.~\eqref{eq:case2_integration_regions}, we use Eq.~\eqref{eq:Energy_dispersion_mus_final} to investigate the relation between the momentum $p_s \to \{p_0, p_c, p_1\}$ and the energies $(E, E_0)$, corresponding to $\cos\theta \in \{0, \cos\theta_c, 1\}$:
\begin{align}
 p_0^2 & = E^2 - E_0^2, \nonumber \\
 \left(\sqrt{p_c^2 + \frac{1}{2}g\sigma|\mu_\Sigma|} - \sqrt{\frac{1}{2}g \sigma |\mu_\Sigma|}\right)^2 & = E^2 - E_0^2, \nonumber\\
 \left(\sqrt{p_1^2 +g^2 \sigma^2} - \frac{1}{2}|\mu_\Sigma|\right)^2 & = E^2. \label{eq:case2_momenta_eqs}
\end{align}

Let us consider first the $>$ and $<$ branches. Here, $E > E_0$ and it can be seen that the above equations allow for two values of $p_c$ and $p_1$:
\begin{gather}
 p_0^2 = E^2 - E_0^2, \quad
 p_{1;\pm}^2 = \left(E \pm \tfrac{1}{2}|\mu_\Sigma|\right)^2 - g^2\sigma^2
 \nonumber\\
 p_{c;\pm}^2 = \sqrt{E^2 - E_0^2} \left(\sqrt{E^2 - E_0^2} \pm \sqrt{2g\sigma |\mu_\Sigma|}\right).
 \label{eq:case2_momenta}
\end{gather}
It is easy to check that, when $E > E_0$, only the solutions corresponding to the upper (positive) sign are meaningful. In the case of $p_c$, we have $p_{c;-} < p_0 < p_{c;+}$, so that $p_{c;-}$ is not part of the integration domain. In the case of $p_1$, the solution originates from $\sqrt{p_1^2 + g^2 \sigma^2} = \frac{1}{2} |\mu_\Sigma| \pm E$. Since $E > \frac{1}{2} |\mu_\Sigma| - g\sigma$, it is clear that $\frac{1}{2} |\mu_\Sigma| - E < g\sigma$, hence $p_{1,-}^2 < 0$ and this is not an acceptable solution. Therefore, the angular integration ranges $\Gamma_\theta^< = [0, \cos\theta_c]$ and $\Gamma_\theta^> = [\cos\theta_c, 1]$ are converted into the momentum integration ranges $p_0 \le p \le p_{c;+}$ and $p_{c;+} \le p \le p_{1;+}$, which can be merged seamlessly. Therefore, the sum $F_q^< + F_q^>$ can be computed by merging the angular integration to $0 \le \cos\theta \le 1$, with the momentum $p_s(\cos\theta)$ continuously and monotonically covering the range between $p_0$ to $p_1$, as given in Eqs.~\eqref{eq:ps01}, with $s \mu_\Sigma \to \frac{1}{2} |\mu_\Sigma|$. Since the above discussion applies equally to $\partial F_q^>/ \partial \sigma + \partial F_q^< / \partial \sigma$, we can directly use the results in Eqs.~\eqref{eq:Fq&dFds_case1_aux}, with the lower end of the energy integration domain changed to $g\sigma - s\mu_\Sigma \to \frac{1}{2}|\mu_\Sigma| - g\sigma$:
\begin{subequations}
\begin{equation}
 F_q^>+F_q^< = -\frac{N_cN_fT}{\pi^2} \int\limits_{\frac{1}{2}|\mu_\Sigma| - g\sigma}^\infty dE\, E p_1^+ \ln (1 + e^{-\beta E}), \label{eq:Fq_case2_outer}
\end{equation}
\begin{multline}
 \frac{\partial F^>_q}{\partial \sigma} +\frac{\partial F^<_q}{\partial \sigma} = \frac{N_cN_f}{\pi^2} g^2 \sigma \int\limits_{\frac{1}{2} |\mu_\Sigma| - g\sigma}^\infty \frac{dE}{e^{\beta E} + 1} \\
 \times \left[p^+_1 - \frac{1}{2} |\mu_\Sigma| \ln \left(\frac{E + \frac{1}{2}|\mu_\Sigma| + p^+_1}{g\sigma}\right)\right],
 \label{eq:dFds_case2_outer}
\end{multline}
\label{eq:Fq&dFds_case2_outer}
\end{subequations}
with $p_1^+ = \sqrt{(E + \frac{1}{2}|\mu_\Sigma|)^2 - g^2 \sigma^2}$.

Moving now to the critical region, we remind that $E$ decreases from $E_0$ to $E_m$ at $p = p_m$, set as a function of $\cos\theta > \cos\theta_c$ by Eq.~\eqref{eq:pm}, then increases back to $E_0$. We now express the integration of the critical region as an integral with respect to $E$ between the minimum value, $E = 0$, and the maximum value, $E_0 = \frac{1}{2}|\mu_\Sigma| - g\sigma$, seeking to establish the integration limits for $\cos\theta$, or equivalently, for $p$. It is not difficult to see that the momentum range is set by the two roots $p_1^\pm$ allowed by Eq.~\eqref{eq:case2_momenta}:
\begin{equation}
 p_1^\pm = \sqrt{\left(\frac{1}{2}|\mu_\Sigma| \pm E\right)^2 - g^2\sigma^2},
\end{equation}
which agrees with the functions $p^s_1$ in Eqs.~\eqref{eq:ps01}, with $s = \pm 1/2$ and $\mu_\Sigma \to |\mu_\Sigma|$.
We thus have
\begin{subequations}
\begin{align}
 F^c_q =& -\frac{2N_cN_f T}{\pi^2 |\mu_\Sigma|}\int\limits_0^{E_0} dE~E~\ln\left(1+e^{-\beta E}\right)\int\limits_{p_1^-}^{p_1^+} \frac{dp\, p}{p_+ p_-} \nonumber\\&\quad\times(p^2 + g^2 \sigma^2 + \tfrac{1}{4}\mu_\Sigma^2 - E^2), \\
 \frac{\partial F^c_q}{\partial \sigma} =&
 \frac{2N_cN_f g^2 \sigma}{\pi^2 |\mu_\Sigma|}\int\limits_0^{E_0} \frac{dE}{e^{\beta E} + 1} \int\limits_{p_1^-}^{p_1^+} \frac{dp\, p}{p_+ p_-} \nonumber\\&\quad\times(p^2 + g^2 \sigma^2 - \tfrac{1}{2} \mu_\Sigma^2 - E^2),
\end{align}
\end{subequations}
with $p^2_\pm = p^2 + (\tfrac{1}{2}|\mu_\Sigma| \pm g\sigma)^2 - E^2$. Switching now to $y = p^2 - (p_1^-)^2$, we arrive at
\begin{subequations}
\begin{align}
 F^c_q &= -\frac{N_cN_f T}{\pi^2 |\mu_\Sigma|}\int\limits_0^{E_0} dE~E~\ln\left(1+e^{-\beta E}\right) \nonumber\\
 &\hspace{1.5cm} \times \int\limits_{0}^Y \frac{dy~y}{\sqrt{(y + y_-)(y + y_+)}},\\
 \frac{\partial F^c_q}{\partial \sigma} &=
 \frac{N_cN_f g^2 \sigma}{\pi^2 |\mu_\Sigma|}\int\limits_0^{E_0} \frac{dE}{e^{\beta E} + 1} \int\limits_{0}^Y \frac{dy(y - |\mu_\Sigma| E)}{\sqrt{(y + y_-)(y + y_+)}},
\end{align}
\end{subequations}
where $y_\pm= |\mu_\Sigma|(\frac{1}{2}|\mu_\Sigma| \pm g\sigma - E)$ and $Y = (p_1^+)^2 - (p_1^-)^2 = 2|\mu_\Sigma| E$. Performing the integration yields
\begin{subequations}
 \begin{align}
 F^c_q &= -\frac{N_cN_f T}{\pi^2} \int\limits_0^{E_0} \frac{dE}{e^{\beta E} + 1} \left(p_1^+ - p_1^- \right)\label{eq:Fq_case2_crit},\\
 \frac{\partial F^c_q}{\partial \sigma} &=
 \frac{N_cN_f g^2 \sigma}{\pi^2} \int\limits_0^{E_0} \frac{dE}{e^{\beta E} + 1} \nonumber\\
 &\times \left[p_1^+ - p_1^-
 %+ |\mu_\Sigma| \ln \frac{\sqrt{y_+} + \sqrt{y_-}}{\sqrt{y_+} - \sqrt{y_-}}
 %- |\mu_\Sigma| \ln \frac{\sqrt{Y_+} + \sqrt{Y_-}}{\sqrt{Y_+} - \sqrt{Y_-}}
 - \frac{1}{2}|\mu_\Sigma| \ln \frac{\frac{1}{2} |\mu_\Sigma| + E + p_1^+}{\frac{1}{2}|\mu_\Sigma| - E + p_1^-}
 \right].
 \label{eq:dFds_case2_crit}
\end{align}
\label{eq:Fq&dFds_case2_crit}
\end{subequations}

We now consider the total free energy, $F_q = F_q^+ + F_q^-$, with the notation introduced at the beginning of this subsection. It is convenient to separate this result as $F_q = F^{\rm reg}_q  + F^{c;+}_q +F^{c;-}_q$, with $F^{\rm reg}_q = F_q^< + F_q^> + F_q^-$ having the same form as in Eq.~\eqref{eq:Fq_case1_aux} of Case 1:
\begin{subequations}
\label{eq:F_case2_total}
\begin{align}
 F^{\rm reg}_q &= -\frac{N_cN_fT}{\pi^2} \sum_{s = \pm \frac{1}{2}} \int\limits_{\frac{1}{2} |\mu_\Sigma| - 2s g\sigma}^\infty dE\, E\, p_1^s \ln (1 + e^{-\beta E}),
 \label{eq:F_case2_reg_final}
\end{align}
with $p_1^s = \sqrt{(E + s|\mu_\Sigma|)^2 - g^2 \sigma^2}$. The contributions $F_q^{c;\pm}$ correspond to the $p_1^\pm$ contributions in Eq.~\eqref{eq:Fq_case2_crit}:
\begin{align}
 F^{c;+}_q &=
 -\frac{N_cN_fT}{\pi^2} \int\limits_0^{\frac{1}{2}|\mu_\Sigma| - g\sigma} dE\, E p_1^+\ln (1 + e^{-\beta E}),
 \label{eq:F_case2_crit_plus} \\
 F^{c;-}_q &=
 \frac{N_cN_fT}{\pi^2} \int\limits_0^{\frac{1}{2}|\mu_\Sigma| - g\sigma} dE\, E p_1^-\ln (1 + e^{-\beta E}).\label{eq:F_case2_crit_minus}
\end{align}
\end{subequations}
where $p_1^\pm = \sqrt{(\frac{1}{2}|\mu_\Sigma| \pm E)^2 - g^2 \sigma^2}$.

Splitting similarly $\partial F_q / \partial \sigma$, we have
\begin{subequations}
\label{eq:dFds_case2_total}
\begin{align}
 \frac{\partial F^{\rm reg}_q}{\partial \sigma} &= \frac{N_cN_f g^2 \sigma}{\pi^2} \sum_{s = \pm\frac{1}{2}} \int\limits_{\frac{1}{2}|\mu_\Sigma| - 2s g\sigma}^\infty \frac{dE}{e^{\beta E} + 1} \nonumber\\
& \times \left[p^s_1 - s|\mu_\Sigma| \ln \left(\frac{E + s|\mu_\Sigma| + p^s_1}{g\sigma}\right)\right],\label{eq:dFds_case2_reg_final}\\
 \frac{\partial F^{c;+}_q}{\partial \sigma} &=
 \frac{N_cN_f g^2 \sigma}{\pi^2} \int\limits_0^{\frac{1}{2} |\mu_\Sigma| - g\sigma} \frac{dE}{e^{\beta E} + 1} \nonumber\\
 & \times \left[p_1^+ - \frac{1}{2}|\mu_\Sigma| \ln \frac{\frac{1}{2}|\mu_\Sigma| + E + p_1^+}{g\sigma}\right], \label{eq:dFds_case2_crit_plus}\\
 \frac{\partial F^{c;-}_q}{\partial \sigma} &=
 -\frac{N_cN_f g^2 \sigma}{\pi^2} \int\limits_0^{\frac{1}{2} |\mu_\Sigma| - g\sigma} \frac{dE}{e^{\beta E} + 1} \nonumber\\
 & \times \left[ p_1^- - \frac{1}{2} |\mu_\Sigma| \ln \frac{\frac{1}{2}|\mu_\Sigma| - E + p_1^-}{g\sigma}\right].\label{eq:dFds_case2_crit_minus}
\end{align}
\end{subequations}
From Eqs.~\eqref{eq:F_case2_total}-\eqref{eq:dFds_case2_total}, we again obtain the Eqs.~\eqref{eq:suitcaseless} given in the main text, by shifting the energies as:
\begin{align}
    E + s|\mu_\Sigma|\to E ~\text{in}~ \eqref{eq:F_case2_reg_final},\eqref{eq:dFds_case2_reg_final}, \nonumber\\
    \tfrac{1}{2}|\mu_\Sigma| + E \to E ~\text{in}~ \eqref{eq:F_case2_crit_plus},\eqref{eq:dFds_case2_crit_plus}, \nonumber\\
    \tfrac{1}{2}|\mu_\Sigma| - E \to E ~\text{in}~ \eqref{eq:F_case2_crit_minus},\eqref{eq:dFds_case2_crit_minus}.\nonumber
\end{align}

In conclusion, we note that even though the two cases (case 1 in Subsec.~\ref{app:sad:case1} and case 2 in Subsec.~\ref{app:sad:case2}) had to be treated differently, suitable manipulations of the corresponding integrals allow the final results to be expressed in a unitary manner, as shown in Eqs.~\eqref{eq:suitcaseless} of the main text.

\section{Small-mass limit of \texorpdfstring{$\delta F_q$}{d Fq}}
\label{app:deltaFq}

To start this section, we refer back to Eq.~\eqref{eq:dFq_def}.
% \begin{equation}
%  \delta F_q = -\frac{1}{2\pi^2} \mu_\Sigma T g^2 \sigma^2 \sum_s s \int_{g\sigma}^\infty \frac{dE}{p} \ln(1 + e^{-\beta|\widetilde{E}_s|}).
% \end{equation}
Absorbing the prefactors within $\delta F_q$ as $\delta F_q' = -\frac{2\pi^2\delta F_q}{\mu_\Sigma T g^2 \sigma^2}$ and performing a derivative with respect to $\mu_\Sigma$ we can rewrite Eq.~\eqref{eq:dFq_def} as
\begin{align}
    4T\frac{\partial\delta F_q'}{\partial\mu_\Sigma} = \sum_s \int_{g\sigma}^\infty \frac{dE}{p} \frac{{\rm sgn}(\widetilde{E}_s)}{e^{\beta|\widetilde{E}_s|} + 1},
\end{align}
where we have used the fact that $s^2=1/4$ is independent of $s$, and have therefore factored it out as a common prefactor on the LHS.
Using Eq.~\eqref{eq:FD_wsignEs_relation} one can then separate $4T\frac{\partial\delta F_q'}{\partial\mu_\Sigma}$ into two parts, that is:
\begin{equation}
4T\frac{\partial\delta F_q'}{\partial\mu_\Sigma} = 4T\frac{\partial\delta F_q'}{\partial\mu_\Sigma}\Bigg|_{T\neq 0} - 4T\frac{\partial\delta F_q'}{\partial\mu_\Sigma}\Bigg|_{T= 0},
\label{eq:deltaFqp_separation}
\end{equation}
with
\begin{align}
    4T\frac{\partial\delta F_q'}{\partial\mu_\Sigma}\Bigg|_{T\neq 0} &= \sum_s \int_{g\sigma}^\infty \frac{dE}{p} \frac{1}{e^{\beta \widetilde{E}_s} + 1},\\
    4T\frac{\partial\delta F_q'}{\partial\mu_\Sigma}\Bigg|_{T = 0} &= \sum_s \int_{g\sigma}^\infty \frac{dE}{p} \theta(-\widetilde{E}_s).
\end{align}
In the limit of small $g\sigma$, one can analytically compute these two contributions, which, to leading order in $\mu_\Sigma/T$, yield the following:
\begin{align}
    4T\frac{\partial\delta F_q'}{\partial\mu_\Sigma}\Bigg|_{T\neq 0} &\simeq -\gamma_E -\ln\frac{g\sigma}{\pi T},\label{eq:deltaFq_Tneq0_integral}\\
    4T\frac{\partial\delta F_q'}{\partial\mu_\Sigma}\Bigg|_{T = 0} &\simeq -\ln\frac{g\sigma}{|\mu_\Sigma|}.\label{eq:deltaFq_Teq0_integral}
\end{align}
Here we have considered the hierarchy $g\sigma < \frac{1}{2}|\mu_\Sigma| < T$ and, to evaluate Eq.~\eqref{eq:deltaFq_Tneq0_integral}, we made use of the integral \cite{kapusta_gale_2006}:
\begin{equation}
    \lim_{y\to 0}\int\limits_0^\infty \frac{dx}{\sqrt{x^2+y^2}}\frac{1}{e^{\sqrt{x^2+y^2}}+1} \simeq -\frac{1}{2}\ln\frac{y}{\pi} -\frac{1}{2}\gamma_E.
\end{equation}
Combining the two contributions from Eqs.~\eqref{eq:deltaFq_Tneq0_integral}-\eqref{eq:deltaFq_Teq0_integral} using Eq.~\eqref{eq:deltaFqp_separation} and integrating with respect to $\mu_\Sigma$ one can eventually obtain
\begin{equation}
    \delta F_q' = \frac{\mu_\Sigma}{4T}\left(1-\gamma_E -\ln\frac{|\mu_\Sigma|}{\pi T}\right),
\end{equation}
in agreement with Eq.~\eqref{eq:smallm_dFq}.

\section{Connection to wave function renormalisation}
\label{app:rg}

In this section, we briefly discuss the connection between the renormalization technique used in this work and the wavefunction renormalization approach employed in Ref.~\cite{Brandt:2025tkg} for the LSM$_q$ model at finite isospin chemical potential $\mu_I$. Employing the approach of Ref.~\cite{Brandt:2025tkg} in our current situation, the field $\sigma$, the parameters $\lambda$ and $v$ and the quark-meson coupling $g$ become renormalization-scale-dependent. Under the wavefunction renormalization scheme, the following combinations are considered to be renormalization-scale invariant:
\begin{equation}
 g_{\rm b} \sigma_{\rm b} = g \sigma = m_q, \quad
 \frac{h_{\rm b}}{g_{\rm b}} = \frac{h_{\rm r}}{g_{\rm r}}, \quad
 \frac{\lambda_{\rm b} v_{\rm b}^2}{g^2_{\rm b}} = \frac{\lambda_{\rm r} v^2_{\rm r}}{g_{\rm r}^2},
\end{equation}
where $b$ and $r$ label the bare and renormalized parameters, respectively. The remaining terms depending on the renormalization energy scale are grouped as follows:
\begin{equation}
 u = \frac{g^4}{\lambda}, \quad
 w = g^2.
\end{equation}

Similar to our approach in Sec.~\ref{sec:vac_ren}, we consider an expression for the meson potential $V_\sigma$ involving bare parameters:
\begin{align}
   V_\sigma^{\rm bare} = &\, \frac{\lambda_{\rm b}}{4}\left(\sigma_{\rm b}^2 -v_{\rm b}^2\right)^2 -h_{\rm b} \sigma_{\rm b} + \ell \mu_\Sigma^2 \sigma_{\rm b}^2 \chi(g_{\rm b} \sigma_{\rm b}) + V_{\rm offset}\nonumber\\
   = &\, \frac{m_q^4}{4u_{\rm b}} - \frac{\lambda v^2}{2g^2} m_q^2 + \frac{\lambda_{\rm b} v_{\rm b}^4}{4} - h\sigma + \frac{\mu_\Sigma^2}{w_{\rm b}}\ell m_q^2 \chi(m_q) \nonumber\\
   & \, + V_{\rm offset},
\end{align}
where $V_{\rm offset}$ is a scale-invariant offset term that will be used to ensure our renormalization conditions given in Eqs.~\eqref{eq:LSM_model_def} and \eqref{eq:scheme2}.
We also added an extra term proportional to $\mu_\Sigma^2$, involving an as-of-yet unspecified function $\chi$, independent of $\mu_\Sigma$ and depending on the meson field only through the scale-invariant combination $g_{\rm b} \sigma_{\rm b} = m_q$. This term involves the parameter $\ell$ introduced in Sec.~\ref{sec:vac_ren:model2} to allow for spin-dependent contributions to the renormalized effective potential. It is akin to the coupling between the isospin chemical potential and the charged pions considered in Ref.~\cite{Brandt:2025tkg}. Such a term is crucial for absorbing the $\mu_I^2$-dependent ultraviolet divergences,
that appear in the case of a finite isospin chemical potential $\mu_I$ considered in Ref.~\cite{Brandt:2025tkg}.
Contrary to Ref.~\cite{Brandt:2025tkg}, the spin charge is not conserved and therefore the spin potential $\mu_\Sigma$ cannot be naturally coupled with other meson degrees of freedom.

To express the relation between the bare and the renormalized, scale-independent parameters, we employ the following notation:
\begin{equation}
 u_{\rm r} = u_{\rm b} Z_u^{-1}, \quad
 w_{\rm r} = w_{\rm b} Z_w^{-1},
 \label{eq:Zdef}
\end{equation}
We then write
\begin{equation}
 V^{\rm bare}_\sigma + F_q^{\rm ZP} = V_\sigma^{\rm ren},
\end{equation}
with $V_\sigma^{\rm ren}$ written in terms of the renormalized parameters as follows
\begin{equation}
 V_\sigma^{\rm ren} = \frac{m_q^4}{4u_{\rm r}} - \frac{\lambda_{\rm r} v_{\rm r}^2}{2g_{\rm r}^2} m_q^2 - h_{\rm r} \sigma_{\rm r} + \frac{\mu_\Sigma^2}{w_{\rm r}} \ell m_q^2 \chi + \Delta V,
 \label{eq:rg_Vren}
\end{equation}
where $\Delta V$ contains only scale-invariant combinations and will be determined shortly.
The fermionic zero-point contribution $F_q^{\rm ZP}$ is given in Eq.~\eqref{eq:IS_def}. It is written with respect to scale-invariant combinations, with the exception of the renormalization scale itself, $M$, appearing in the logarithmically-divergent terms in $I_1$ and $I_2$, given in Eqs.~\eqref{eq:I1_reg} and \eqref{eq:I2I3_reg}, respectively. We further identify the dimensionless regularization parameter $\epsilon$ with the mass scale via
\begin{equation}
 \frac{1}{\epsilon} = 2\ln \frac{\Lambda}{M},
\end{equation}
where $\Lambda$ can be interpreted as a momentum UV cutoff, while the factor $2$ is conventional.
We therefore find
\begin{align}
 Z_u &= 1 + \frac{N_c N_f u_{\rm b}}{4\pi^2} \left(\frac{3}{2} - \gamma_E + \ln \frac{4\pi \Lambda^2}{M^2}\right), \nonumber\\
 Z_w &= 1 + \frac{N_c N_f w_{\rm b}}{16\pi^2 \ell \chi(m_q)} \left(-\gamma_E + \ln \frac{4\pi \Lambda^2}{M^2}\right).
\end{align}
We see that $\Delta V$ is given by
\begin{multline}
 \Delta V = \frac{\lambda_{\rm b} v_{\rm b}^4}{4} +
 \frac{N_c N_f m_q^2}{16\pi^2} (m_q^2 + \mu_\Sigma^2) \ln \frac{M^2}{m_q^2} \\
 - 2N_c N_f (\mu_\Sigma^4 I_3 + \delta I_\Sigma) + V_{\rm offset}.
\end{multline}

We are now ready to test for scale invariance. Considering $V^{\rm ren}_{\mathcal{M}}$ a function of $M$ explicitly, as well as implicitly, through the running parameters $u_{\rm r} = u_{\rm b} Z_u^{-1}$ and $w_{\rm r} = w_{\rm b} Z_w^{-1}$, the scale invariance demands
\begin{equation}
 \left(M\frac{\partial}{\partial M} + \beta_u \frac{\partial}{\partial u_{\rm r}} + \beta_w \frac{\partial}{\partial w_{\rm r}}\right) V^{\rm ren}_\sigma = 0.
 \label{eq:RG_dVdM}
\end{equation}
The beta functions are defined as
\begin{align}
 \beta_u = M\frac{du_{\rm r}}{dM} &= \frac{N_c N_f}{2\pi^2} u_{\rm r}^2, \nonumber\\
 \beta_w = M\frac{dw_{\rm r}}{dM} &= \frac{N_c N_f}{8\pi^2 \ell \chi} w_{\rm r}^2\,.
 \label{eq:RG_beta}
\end{align}
To see that Eq.~\eqref{eq:RG_dVdM} is satisfied, we compute below the partial derivatives of $V^{\rm ren}_{\mathcal{M}}$ with respect to $M$, $u_{\rm r}$ and $w_{\rm r}$:
\begin{gather}
 M\frac{\partial V^{\rm ren}_\sigma}{\partial M} = \frac{N_c N_f m_q^2}{8\pi^2} (m_q^2 + \mu_\Sigma^2), \nonumber\\
 \frac{\partial V^{\rm ren}_\sigma}{\partial u_{\rm r}} = -\frac{m_q^4}{4u_{\rm r}^2}, \quad
 \frac{\partial V^{\rm ren}_\sigma}{\partial w_{\rm r}} = -\frac{\mu_\Sigma^2 m_q^2}{w_{\rm r}^2} \ell \chi.
 \label{eq:RG_partialV}
\end{gather}
Substituting Eqs.~\eqref{eq:RG_beta} and \eqref{eq:RG_partialV} into Eq.~\eqref{eq:RG_dVdM} shows that $V^{\rm ren}_\sigma$ is indeed scale invariant. We can thus evaluate all running parameters of $V^{\rm ren}_\sigma$ at the scale $M = M_0 = g f_\pi$. This can be performed by integrating the beta functions in Eq.~\eqref{eq:RG_beta}:
\begin{align}
 \frac{1}{u_{\rm r}} = &\, \frac{1}{u_0} + \frac{N_c N_f}{2\pi^2} \ln \frac{M_0}{M}\,, \\
 \frac{1}{w_{\rm r}} = &\, \frac{1}{w_0} + \frac{N_c N_f}{8\pi^2 \ell \chi} \ln \frac{M_0}{M}\,,
\end{align}
leading to
\begin{align}
	 V^{\rm ren}_\sigma = &\, \frac{m_q^4}{4u_0} - \frac{\lambda_0 v_0^2}{2g_0^2} m_q^2 - h_0 \sigma_0 + \frac{\mu_\Sigma^2}{w_0} \ell m_q^2 \chi + \frac{\lambda_{\rm b} v_{\rm b}^4}{4} \nonumber \\
 &\, + \frac{N_c N_f m_q^2}{16\pi^2} (m_q^2 + \mu_\Sigma^2) \ln \frac{M_0^2}{m_q^2} \nonumber\\
 &\, - 2N_c N_f (\mu_\Sigma^4 I_3 + \delta I_\Sigma) + V_{\rm offset}.
 \label{eq:Vren_aux}
\end{align}

The quantities with a zero subscript: $\lambda_0$, $v_0$, $g_0$, $h_0$, $u_0 = g_0^4 / \lambda_0$ and $w_0 = g_0^2$ reduce to the model parameters in Eq.~\eqref{eq:LSM_model_def} and will therefore be shown without the $0$ subscript. In order to ensure Eq.~\eqref{eq:LSM_model_def}, $V_{\rm offset}$  must be taken as
\begin{align}
	 V_{\rm offset} = &\, -\frac{\lambda f_\pi^4}{4}   + \frac{\lambda v^2 f_\pi^2}{2} - \frac{\lambda_{\rm b} v_{\rm b}^4}{4} + h f_\pi \nonumber \\
    &\,  - \frac{N_c N_f}{16\pi^2} g^4 \sigma^2 \ln \frac{f_\pi^2}{\sigma^2} + \delta V_{\rm offset},
\end{align}
with $\lim_{\mu_\Sigma \to 0} \delta V_{\rm offset}= 0$. Thus, Eq.~\eqref{eq:Vren_aux} takes the form
\begin{multline}
 V^{\rm ren}_\sigma = \frac{\lambda}{4}[(\sigma^2 - v^2)^2 - (f_\pi^2 - v^2)^2] - h(\sigma - f_\pi) \\
 + \mu_\Sigma^2 \sigma^2 \ell \bar{\chi}
 - 2N_c N_f (\mu_\Sigma^4 I_3 + \delta I_\Sigma) + \delta V_{\rm offset},
\end{multline}
where we introduced the notation
\begin{equation}
 \bar{\chi} = \chi + \frac{N_c N_f g^2}{16\pi^2 \ell} \ln \frac{f_\pi^2}{\sigma^2}.
\end{equation}

The conditions in Eq.~\eqref{eq:scheme2} entail
\begin{align} \label{eq:RG:model2:relations}
 \left[\ell \mu_\Sigma^2 \frac{\partial (\sigma^2 \bar{\chi})}{\partial \sigma} + \frac{\partial}{\partial \sigma} (\delta V_{\rm offset} - 2 N_c N_f \delta I_\Sigma) \right]_{\sigma = f_\pi} \hspace{-15pt} &= 0,\nonumber\\
 \left[\ell \mu_\Sigma^2 \frac{\partial^2 (\sigma^2 \bar{\chi})}{\partial \sigma^2} + \frac{\partial^2}{\partial \sigma^2} (\delta V_{\rm offset} - 2 N_c N_f \delta I_\Sigma) \right]_{\sigma = f_\pi} \hspace{-15pt} &= 0.
\end{align}
While there is no {\it a priori} scheme to satisfy the above conditions uniquely, we restore the scheme in Sec.~\ref{sec:vac_ren:model2} by choosing:
\begin{align}
 \bar{\chi} &= \frac{N_c N_f}{32\pi^2 f_\pi^2} \left(m_q^2 - \frac{M_0^4}{m_q^2} + 2 M_0^2 \ln \frac{M_0^2}{m_q^2}\right), \nonumber\\
 &= \frac{N_c N_f g^2}{32\pi^2 f_\pi^2} \left(\sigma^2 - \frac{f_\pi^4}{\sigma^2} + 2 f_\pi^2 \ln \frac{f_\pi^2}{\sigma^2}\right), \nonumber\\
 \delta V_{\rm offset} &= 2N_c N_f (\mu_\Sigma^4 I_3 + \delta I_\Sigma).
 \label{eq:RG:model2}
\end{align}
When $\sigma = f_\pi$, it can be seen that $\bar{\chi}$ satisfies:
\begin{equation}
 \bar{\chi} \to 0, \quad
 \frac{\partial \bar{\chi}}{\partial \sigma} \to 0, \quad
 \frac{\partial^2 \bar{\chi}}{\partial \sigma^2} \to 0,
\end{equation}
thereby ensuring the relations in Eq.~\eqref{eq:RG:model2:relations}. We therefore conclude that the effective potential considered in Eq.~\eqref{eq:LSM_mus_model2} is renormalization-scale invariant.

% \paragraph{Inclusion of the remainder term :}
% One simplest way to include the remainder term satisfying all the renormalization conditions and keeping the vacuum pressure 0 at $\sigma=f_\pi$ is to have,
% \begin{align}
%     V_\mathcal{M}&= V_\sigma^{\mathrm{ren}}+\theta(\mu_\Sigma-2 m_q) V_{\mathrm{rem}}+\nonumber\\& -\theta(\mu_\Sigma-2M_0)\left[\left(\frac{m_q}{M_0}-1\right)f_\pi \frac{\partial V_{rem}}{\partial \sigma}\big{|}_{\sigma=f_\pi}+\right.\nonumber\\&\left.\frac{\left(\frac{m_q}{M_0}-1\right)^2}{2}f_\pi^2 \frac{\partial^2 V_{\mathrm{rem}}}{\partial \sigma^2}\Big{|}_{\sigma=f_\pi} + V_{\mathrm{rem}}\Big{|}_{\sigma=f_\pi} \right]
% \end{align}
% where $V_\sigma^{\mathrm{ren}}$ is given in Eq.\eqref{eq:LSM_mus_model2} , and $V_{\mathrm{rem}}= -2N_cN_f\delta I_{\Sigma}$, with $\delta I_\Sigma$ given in Eq.\eqref{eq:dIS_result}.
% However, inclusion of these corrections do not influence the phase transition properties and the phase diagram until $\mu_\Sigma$ becomes significantly high, hence the comparison with the lattice result remains unaffected. at zero temperature on the other hand where these terms have the most dominant effect they tend to cancel the effect of the $\mathcal{O}(\mu_\Sigma^2)$ term and force the system towards confined phase.

\bibliography{ref}

\end{document}